\documentclass[prd,showpacs,superscriptaddress,preprintnumbers,amsmath,nofootinbib]{revtex4}
 \usepackage[dvips,final]{graphicx}
  \usepackage{amssymb,amsmath,epsfig,bm,pifont}
   \usepackage{extarrows}
\usepackage{color}

   \definecolor{DarkGreen}{rgb}{0.04,0.5,0.1}

\usepackage{relsize}
\newcommand{\babar}{{\mbox{\slshape B\kern-0.1em{\smaller A}\kern-0.1em
            B\kern-0.1em{\smaller A\kern-0.2em R}}}
\def\MSbar{\relax\ifmmode\overline                        %%%%%%%%%
            {\rm MS}\else{$\overline{\rm MS}${ }}\fi}     %%%%%%%%%
           }                                              %%%%%%%%%
\newcommand{\Ds}{\displaystyle}                           %%%%%%%%%
\def\MSbar{\relax\ifmmode\overline                        %%%%%%%%%
            {\rm MS}\else{$\overline{\rm MS}${ }}\fi}     %%%%%%%%%
\def\1{\hbox{{1}\kern-.25em\hbox{l}}}
 \date{\today}
 \preprint{\hbox{RUB-TPII-02/2018}}
\def\Im{\relax{\textbf{Im}{}}}                            %%%%%%%%%
\def\Re{\relax{\textbf{Re}{}}}                            %%%%%%%%%
\def\be{\begin{equation}}
\def\ee{\end{equation}}
\def\bea{\begin{eqnarray}}
\def\eea{\end{eqnarray}}
\def\bear{\begin{array}}
\def\eear{\end{array}}

\newcommand{\A}{{\mathcal{A}}}
\newcommand{\M}{{\mathfrak{A}}}
\newcommand{\I}{{\mathcal{I}}}

\begin{document}
\title{Calculation of the pion-photon transition form factor using
       dispersion relations and renormalization-group
       summation\footnote{Dedicated to the memory of D.~V.~Shirkov}}

\author{C\'esar Ayala}
\email{cesar.ayala@usm.cl}
\affiliation{Department of Physics,
             Universidad T\'ecnica Federico Santa Mar\'ia,
             Casilla 110-V, Valpara\'iso, Chile\\}

\author{S.~V.~Mikhailov}
\email{mikhs@theor.jinr.ru}
\affiliation{Bogoliubov Laboratory of Theoretical Physics, JINR,
             141980 Dubna, Russia\\}

\author{N.~G.~Stefanis}
\email{stefanis@tp2.ruhr-uni-bochum.de}
\affiliation{Institut f\"{u}r Theoretische Physik II,
             Ruhr-Universit\"{a}t Bochum,
             D-44780 Bochum, Germany\\}

\date{\today}

\begin{abstract}
We consider the lightcone sum-rule description of the pion-photon
transition form factor, based on dispersion relations, in combination
with the renormalization group of QCD, in terms of the formal solution
of the Efremov-Radyushkin-Brodsky-Lepage evolution equation, and show
that the emerging scheme amounts to a certain version of Fractional
Analytic Perturbation Theory (FAPT).
In order to ensure the correct asymptotic behavior of the considered
physical quantity, this modified FAPT version has to be supplemented by
process-specific boundary conditions---in contrast to the standard one.
However, it provides the advantage of significantly improving the
inclusion of radiative corrections in the low-momentum regime of QCD
perturbation theory using renormalization-group summation.
\end{abstract}
%               11.10.Hi Renormalization group evolution of parameters
%               12.38.Bx Perturbative calculations
%               12.38.Cy Summation of perturbation theory
%               12.38.Lg Other nonperturbative calculations
\maketitle

\section{Introduction}
\label{sec:intro}
The description of hard exclusive hadronic processes in QCD is
difficult because it must account for typical nonperturbative phenomena
like the hadron binding dynamics and/or long-distance effects pertaining
to soft contributions that cannot be assessed by means of perturbative
Quantum Chromodynamics (pQCD).

Consider for example the pion-photon transition form factor for two highly
virtual photons describing the reaction
$\gamma^*(-Q^2)\gamma^*(-q^2) \to \pi^0$
by assuming that
$Q^2, q^2 \gg m^2_\rho$.
Applying factorization, the pion-photon transition form factor (TFF) is
given by the following correlation function written in a generic
convolution form as follows
\begin{eqnarray}
  F^{\gamma^*\gamma^*\pi^0}(Q^2,q^2,\mu^2)
\sim && \!\!\!
   T^{(2)}(Q^2,q^2,\mu^2;x)\underset{x}\otimes \varphi_{\pi}^{(2)}(x,\mu^2)
   + T^{(4)}(Q^2,q^2,\mu^2;x)\underset{x}\otimes \varphi_{\pi}^{(4)}(x,\mu^2)
\nonumber \\
&&  \!\!\! + ~\text{higher inverse-power corrections} ,
\label{eq:TFF-convol}
\end{eqnarray}
%Eq (1)
where
$\underset{x}\otimes \equiv \int_0^1 dx$
and the superscript $(n)$ is the twist label.
For simplicity, we have adopted the default scale setting, i.e.,
$\mu_\text{F}=\mu_\text{R}=\mu$, where the abbreviations refer to the
factorization and renormalization scales, respectively.

A useful calculational scheme to implement a consistent factorization of
short-distance dynamics, amenable to QCD perturbation theory via hard-gluon
exchanges, from long-distance phenomena, encoded in nonperturbative hadron
distribution amplitudes based on the lightcone operator product expansion
(OPE), is provided by lightcone sum rules (LCSR)s
\cite{Balitsky:1989ry,Khodjamirian:1997tk}.
In this scheme, correlation function (\ref{eq:TFF-convol})
can be cast in the form of a dispersion relation in terms of the
large photon virtuality $Q^2$ to obtain a LCSR.
This dispersive conceptual picture of exclusive hadronic processes
will be a key issue in the present investigation.

The pion-photon TFF
represents a prototypical example of such a process and
provides valuable information on the quark
structure of the pion in terms of its leading twist two
(and subleading twist four) distribution amplitudes (DA)s
$\varphi_{\pi}^{(2,4)}(x)$.
Moreover, it can be measured in single-tag experiments.
A classification of various theoretical predictions in comparison
with the available data
\cite{Behrend:1990sr,Gronberg:1997fj,Aubert:2009mc,Uehara:2012ag}
can be found in \cite{Bakulev:2012nh}.
In contrast, the pion DA is not directly measurable but has to be
inferred from the data or be constructed from nonperturbative models.
In most theoretical analyses, it is reversed engineered from its
(first few) moments \cite{Chernyak:1983ej}
(see also \cite{Stefanis:1999wy})
\begin{equation}
  \langle \xi^{N} \rangle_{\pi}
\equiv
  \int_{0}^{1} dx (2x-1)^{N} \varphi_{\pi}^{(2)}(x,\mu^2) \, ,
\label{eq:DA-moments}
\end{equation}
%Eq (2)
where $N=2,4,\ldots$,
$\xi=2x-1=x-\bar{x}$, $\bar{x}=1-x$,
with $x$ being the longitudinal momentum fraction carried
by the valence quark in the pion.
Until now only the second moment $\langle \xi^{2} \rangle_{\pi}$
has been measured on the lattice,
yielding diverging values
\cite{Braun:2006dg,Braun:2015axa,Arthur:2010xf,Segovia:2013eca,Bali:2017ude,Bali:2018spj},
while there are not enough experimental data to constrain the moments
above $N=6$.
For a discussion of these techniques and comparison of various types
of pion DAs with the lattice estimates, see \cite{Bakulev:2011rp}.
More advanced theoretical aspects of the light-meson DAs have been
considered in \cite{Stefanis:2014nla,Stefanis:2014yha,Stefanis:2015qha}
with arguments based partly on QCD sum rules (SR)s
with nonlocal condensates
\cite{Mikhailov:1986be,Mikhailov:1988nz,Mikhailiov:1989mk,Mikhailov:1991pt}.
There are also alternative computational methods, for instance,
Dyson-Schwinger equations \cite{Chang:2013pq,Raya:2015gva},
AdS/QCD \cite{Brodsky:2011yv,Brodsky:2011xx}, etc.

On the other hand, we have in our hands the very powerful method of the
QCD renormalization group (RG) that tells us how QCD properties are
related to each other at different momentum scales (notably,
the strong coupling and dynamical quantities like various parton
distribution functions via their anomalous dimensions).
A serious problem one encounters when applying QCD perturbation theory
is that the running coupling $\alpha_s(Q^2)$ increases logarithmically
at low $Q^2$ so that the validity of the expansion rapidly deteriorates
when $Q^2 \sim \Lambda_\text{QCD}^2$, giving rise to the Landau
singularity.
This affects the proper inclusion of higher-order radiative corrections and
the determination of the optimal choice of the renormalization-scale
setting procedure considerably
\cite{Stefanis:1998dg,Stefanis:2000vd,Bakulev:2004cu}.

To avoid this problem, analytic versions of the power-series
expansion in $\alpha_s$ (better say, non-power expansions) have been
proposed by various authors, e.g.,
\cite{Grunberg:1982fw,Milton:1996fc,Milton:1997us,Shirkov:1997wi,%
Grunberg:1997ud,Nesterenko:1999np,Nesterenko:2003xb,Magradze:1999um}
(see \cite{Shirkov:2006gv} for a review and further references
while more recent developments are discussed, for instance, in
\cite{Cvetic:2011ym,Nesterenko:2017wpb}).
Such schemes make use of dispersion relations in the spacelike and the
timelike regions in order to implement causality while preserving the
RG properties---see \cite{Nesterenko:2016pmx} for a broad review of such
methods.
For our analysis below, we mention explicitly the
Analytic Perturbation Theory (APT) \cite{Shirkov:1997wi} and its
generalization to any real power of the coupling constant,
described by Fractional APT (FAPT) \cite{Bakulev:2005gw}.

As already mentioned with respect to the pion-photon TFF,
LCSRs enable the calculation of various physical quantities
on the basis of dispersion relations.
Their main ingredient is a spectral density that can be calculated in
terms of the hard-scattering amplitude for the quark-gluon subprocesses
order-by-order in QCD perturbation theory.
It appears therefore natural to investigate the LCSR approach in
conjunction with the renormalization group and see how the
LCSR dispersion representation can match a RG-improved
perturbative expansion.
To achieve this goal, we will have to invent a particular
version of FAPT that employs process-dependent boundary conditions
on the behavior of the coupling in the deep infrared (IR) regime in order to ensure compliance
with the QCD asymptotics.
The strategy is to develop a scheme with the advantage of including
the RG series of radiative corrections to the TFF at once.

The rest of the paper is organized as follows.
In the next section (Sec.\ \ref{sec:tff-rg-sum}) we
expose the key idea of RG improvement by expressing the TFF as
a convolution of hard-scattering amplitudes in QCD perturbation theory
with the twist-two pion distribution amplitude.
Section \ref{sec:lcsr-radiative-corr} deals with the inclusion of
radiative corrections into the LCSR using a dispersive
representation in conjunction with the RG approach.
To realize this goal, we have to ``calibrate'' the behavior of the FAPT
analytic coupling at $Q^2=0$ in such a way as to ensure the correct
asymptotic behavior of the TFF at $Q^2\to \infty$.
Drawing on these ideas, we develop an extended version of FAPT---originally developed in
\cite{Bakulev:2005gw,Bakulev:2006ex,Bakulev:2010gm}
with recourse to \cite{Karanikas:2001cs}
(see also \cite{Stefanis:1998dg,Stefanis:2000vd,Bakulev:2004cu,Bakulev:2005fp}
and \cite{Bakulev:2008td,Stefanis:2009kv} for reviews)---by augmenting this
perturbation theory with a new analytic charge $\mathcal{I}_n$ that amends
the conflict between the FAPT analytic couplings at $Q^2=0$ and the asymptotic
behavior of the TFF following from QCD.
In Sec.\ \ref{sec:mesonic-FAPT} we discuss how the soft, i.e., quasi-real
photon in the TFF, relevant for single-tag experiments,
can be accommodated within the new FAPT framework.
Phenomenological implications of our theoretical scheme are discussed
in Sec.\ \ref{sec:phenomenology}.
Our conclusions are given in Sec.\ \ref{sec:concl}, while some important
technical issues are treated in three Appendices.

\section{TFF with RG improvement}
\label{sec:tff-rg-sum}
In this section we consider the pion-photon TFF in convolution form
\cite{Bakulev:2002uc,Mikhailov:2009kf,Mikhailov:2016klg}
and perform a RG summation with the aim to obtain a LCSR in terms of an
improved dispersion relation (see Sec.\ \ref{sec:lcsr-radiative-corr}).
The key idea of our procedure here and below is the following.
We combine \textit{causality}, encoded in the dispersion relations of LCSRs,
with \textit{RG invariance}, which induces \textit{analyticity} of the
perturbative expansion in order to transfer the power-series expansion
of the pion-photon TFF in terms of the usual QCD coupling (and its powers),
exhibiting ghost singularities, into a functional expansion over singularity-free
``calibrated'' analytic couplings that preserve the UV asymptotics of this observable.
Let us now enter the formal description of this task.

At the twist-two (tw$=2$) level, the amplitude for the
hard process
$\gamma^*(-Q^2)\gamma^*(-q^2) \to \pi^0$,
where the two photon virtualities are subject to the condition
$Q^2\gg m^2_\rho$ and $q^2 > m^2_\rho$, can be written in the general
form (referring for the partial cases of the TFF to
\cite{Melic:2002ij,Mikhailov:2009kf})
\begin{eqnarray}
  F^\text{(tw=2)}(Q^2,q^2)
&=&
  N_\text{T} T_0(y) \underset{y}{\otimes}
                                         \left\{\vphantom{\int_{a_s}^{\bar{a}_s(y)}}\!\left[\1
                                                        +\bar{a}_s(y)\mathcal{T}^{(1)}(y,x)
                                                        +\bar{a}_s^2(y)\mathcal{T}^{(2)}(y,x)
                                                        + \ldots
                                                \right]
                    \underset{x}{\otimes} \right.
\nonumber \\
&&
                                         \left.
                                         \exp\left[-\int_{a_s}^{\bar{a}_s(y)} d\alpha \,
                                                    \frac{V(\alpha;x,z)}{\beta(\alpha)}
              \right]
                                        \right\}
                   \underset{z}\otimes \varphi_{\pi}^{(2)}(z,\mu^2) \, ,
              \label{eq:Tfin}
\end{eqnarray}
%Eq (3)
where we have incorporated the general solution of the RG equation for
the QCD charge pertaining to the $\beta$-function,
$\beta(a_s)= -a_s^2(\beta_0+ a_s\beta_1+\ldots)$, by
$\bar{a}_s(y) \equiv \bar{a}_s(q^2\bar{y}+Q^2y)$.
Moreover, we have introduced the coupling parameter
$a_s(\mu^2)\equiv \alpha_s(\mu^2)/4\pi$
with $a_s \equiv a_s(\mu^2=\mu^2_\text{F}=\mu^2_\text{R})$
and the color factor
$\Ds N_\text{T}=\sqrt{2}f_\pi/3$, whereas the pion
decay constant has the value $f_\pi=132$~MeV.
The other elements of the above equation have the following meaning:
$T_0(y)\equiv T_0(Q^2,q^2; y)=1/(q^2\bar{y}+Q^2y)$ is the Born term
of the hard-scattering amplitude,
$\1= \delta(x-y)$, and
$\mathcal{T}^{(i)}$ is the coefficient function of the quark-gluon
subprocess at the loop order $i$, where
$V(a_s)= a_s V_0+a_s^2 V_1+ \ldots$
denotes the evolution kernel related to the perturbative expansion of
the Efremov-Radyushkin-Brodsky-Lepage (ERBL)
evolution equation \cite{Efremov:1978rn,Lepage:1980fj}.
For convenience later on, we have also introduced the abbreviation
$Q(y)\equiv q^2\bar{y}+Q^2y$
that represents the effective virtuality of the ``hand-bag'' diagrams.

The integration over $y$ in (\ref{eq:Tfin}) is possible for large
enough values of $q^2$, at least for $q^2 > \Lambda^2_\text{QCD}$,
so that one remains within the allowed range of pQCD.
Strictly speaking, one has to employ
$q^2 \sim \mu^2_\text{F} \gg \Lambda^2_\text{QCD}$
in order to ensure that the calculations are performed
within the domain of applicability of the factorization approach of pQCD.
At the one-loop level, the next-to-leading-order (NLO) coefficient
function is $\mathcal{T}^{(1)}$ and Eq.\ (\ref{eq:Tfin}) reduces
in the basis of the Gegenbauer harmonics
$\psi_n(x)=6x\bar{x}C^{3/2}_n(x-\bar{x})$
to
\begin{subequations}
\label{eq:RGsum}
\begin{eqnarray}
  F_{n}^\text{(tw=2)}(Q^2,q^2)\xlongrightarrow{\text{1-loop}} F_{(1l)n}^\text{(tw=2)}
&=&N_\text{T}T_0(y)
\underset{y}{\otimes}
                     \left\{
                            \left[\1+ \bar{a}_s(y)\mathcal{T}^{(1)}(y,x)\right]
                            \exp\left[\frac{1}{2}\int_{a_s}^{\bar{a}_s(y)}
                                      \frac{d\alpha}{\alpha} \frac{\gamma_0(n)}{\beta_0} \right]
                     \right\}
\underset{x}\otimes
                     \psi_n(x)
\label{eq:T1c}\\
&=&
  N_\text{T}T_0(y)
\underset{y}{\otimes}
                    \left\{
                           \left[\1+ \bar{a}_s(y)\mathcal{T}^{(1)}(y,x)\right]
                           \left(\frac{\bar{a}_s(y)}{a_s(\mu^2)} \right)^{\nu_n}
                    \right\}
\underset{x}\otimes
                    \psi_n(x) \, .
\label{eq:T1d}
\end{eqnarray}
\end{subequations}
%Eqs. (4a), (4b)
The above equation follows from the relations
\begin{equation}
  V(\alpha;y,z) \to \alpha\cdot V_0(y,z);~~~
  V_0(y,z)\otimes\psi_n(z)=-\frac{1}{2}\gamma_0(n)\psi_n(y);~~~
  \beta(\alpha) \to -a_s^2\beta_0 \, ,
\label{eq:various}
\end{equation}
%Eq (5)
where $a_s\gamma_{0}(n)$ denotes the one-loop anomalous dimension of
the corresponding composite operator of leading twist
with $\displaystyle\nu_n=\frac{1}{2}\frac{\gamma_0(n)}{\beta_0}$.
Finally, the function $\psi_n$ represents the $n$th-harmonic
contribution in the conformal expansion of
$\varphi_{\pi}^{(2)}(x,\mu^2)$, i.e.,
\begin{equation}
  \varphi_{\pi}^{(2)}(x,\mu^{2})
=
  \psi_{0}(x)
  + \sum_{n=2,4, \ldots}^{\infty} a_{n}(\mu^{2}) \psi_{n}(x)\, .
\label{eq:gegen-exp}
\end{equation}
%Eq (6)
Because the moments $\langle \xi^{N} \rangle_{\pi}$
($N=2,4, \ldots$)
and the conformal coefficients $a_n$ are interrelated,
once the moments of the DA have been extracted, one can compute
a subset of $a_n$ within a margin of theoretical uncertainties at
the same normalization scale \cite{Bakulev:2004mc}.
Employing expansion (\ref{eq:gegen-exp}), the leading twist TFF in
Eq.\ (\ref{eq:Tfin}) reads
\begin{equation}
  F^\text{(tw=2)}(Q^2,q^2)
=
  F_{0}^\text{(tw=2)}(Q^2,q^2)
  + \sum_{n=2,4, \ldots}^{\infty} a_{n}(\mu^{2}) F_{n}^\text{(tw=2)}(Q^2,q^2)\, .
\label{eq:FFgegen-exp}
\end{equation}
%Eq (7)
By virtue of $\psi_{0}(x)=\varphi_{\pi}^\text{asy}=6x\bar{x}$,
$\varphi_{\pi}^{(2)}(x)$ practically reduces to the set of
$\psi_{n}(x)$, while the ERBL evolution in (\ref{eq:RGsum})
is governed by the powers $\displaystyle\nu_n$.

Thus, the reduced formula (\ref{eq:T1d}) accumulates the one-loop
RG running of $\bar{a}_s$ and also the analogous one entering the
common ERBL factor to all orders of the perturbative expansion.
The contribution of the zeroth-order harmonic assumes the simplest
form $\gamma_i(0)=0$, $\nu_0=0$ due to the current conservation
$j_{5\mu}=\bar{q}\gamma_5\gamma_\mu q$.
Therefore, in this case, expression (\ref{eq:T1d}) finally reduces to
\begin{equation}
  F_{n=0}^\text{(tw=2)}(Q^2,q^2)
=
  N_\text{T}T_0(y)
\underset{y}{\otimes}
  \left[\1+ \bar{a}_s(y)\mathcal{T}^{(1)}(y,x)
  \right]
\underset{x}\otimes
  \psi_0(x)\, .
\label{eq:T1dn0}
\end{equation}
%Eq (8)

Expanding $\bar{a}_s(y)$ and the ERBL factors in (\ref{eq:T1d}), one
recovers the results stemming from the radiative corrections to
the TFF at the next-to-next-to-leading order (NNLO) level
\cite{Mikhailov:2016klg}.
Indeed, within the framework of fixed-order perturbation theory (FOPT),
the hard-scattering amplitudes have the following structure
\begin{equation}
  F_\text{FOPT}^\text{(tw=2)}(Q^2,q^2)
=
  N_\text{T}
            \left(
                  T_{\rm LO}+ a_s T_{\rm NLO} + a_s^2 T_{\rm NNLO}+\ldots
            \right)
\otimes
  \varphi_{\pi}^{(2)} \, .
\label{eq:F2-FOPT}
\end{equation}
%Eq (9)
The various radiative corrections are given by
\begin{subequations}
\label{eq:hard-scat-series}
\begin{eqnarray}
  T_{\rm LO} \, ,
& = &
  a_s^0~ T_0(x) \\
  a_s T_{\rm NLO}
& = &
  a_s^1~ T_0(y) \otimes \left[\mathcal{T}^{(1)}+\underline{L~  V_{0}}
              \right](y,x) \, ,
\label{eq:NLO}
\\
  a_s^2 T_{\rm NNLO}
& = &
  a_s^2~T_0(y) \otimes \left[
                     {\cal T}^{(2)}
                     - \underline{L~{\cal T}^{(1)} \beta_{0}  }+\underline{L~{\cal T}^{(1)}\otimes V_{0}}
                     - \underline{\frac{L^2}{2}~ \beta_{0} V_{0}}
                     + \underline{\frac{L^2}{2}~ V_{0}^{}\otimes V_{0}}
            + \underline{\underline{L~ V_{1}}} \right](y,x) \, ,
\label{eq:NNLO}
\end{eqnarray}
\end{subequations}
%Eq (10a), (10b), (10c)
where $L=L(y)=\ln\left[(q^2\bar{y}+Q^2y )/\mu^2\right]$.
The underlined terms in Eq.\ (\ref{eq:hard-scat-series})
pertain to the running coupling $\bar{a}_s(y)$ and the common ERBL factor
in Eq.\ (\ref{eq:RGsum}).
The remaining plain terms represent the one-loop, ${\cal T}^{(1)}$,
and the two-loop, ${\cal T}^{(2)}$, corrections, respectively---cf.\
Eq.\ (\ref{eq:Tfin}), first line.
Finally, the double underlined term in Eq.\ (\ref{eq:NNLO}) marks
the beginning of the next ``tower'' of two-loop corrections to the
common term $\bar{a}_s(y)$ and the ERBL factor in the general expression
given by Eq.\ (\ref{eq:Tfin}).
The explicit expressions for ${\cal T}^{(1)}$ and $V_{0}$ are presented
in Appendix \ref{App:A}, while the equations for $V_{1}$ and the elements
of ${\cal T}^{(2)}$ and related references can be found in Appendix A in
\cite{Mikhailov:2016klg}.

Let us emphasize at this point that one cannot directly use the formulas
given by (\ref{eq:RGsum}) for small $q^2$ --- even if $Q^2$ is large.
The reason is that for $q^2 < \mu_\text{F}^2$,
these expressions run out of their applicability domain allowed by
the ``factorization conditions'' mentioned before in this section.
Indeed, the scale argument $q^2\bar{y}+Q^2y$ becomes for $y\to 0$
smaller than $\mu_\text{F}^2$ and hence unprotected.
For this reason, we do not use the complete result of the RG summation
at small $q^2$, but employ instead the FOPT one, see, for instance,
Sec.\ 2 in Ref.\ \cite{Mikhailov:2016klg}.
The situation changes completely when one applies the results expressed
via Eqs.\ (\ref{eq:RGsum}), (\ref{eq:hard-scat-series}) to a dispersion relation
as we are now going to show.

\section{Radiative corrections to the TFF using a dispersive representation}
\label{sec:lcsr-radiative-corr}
The aim in this section is to discuss the radiative corrections
to the pion-photon TFF using the dispersion-relation
representation that forms the basis of the sum rules on the
lightcone \cite{Khodjamirian:1997tk} (see also \cite{Agaev:2010aq}
for a more recent exposition of the method).

In this formalism, the TFF satisfies the following dispersion relation
\begin{equation}
  F_{\gamma^*\gamma\to\pi^0}^\text{LCSR}(Q^2,q^2)
=
  N_\text{T} \int_{0}^\infty\frac{\rho(Q^2,s)}{s+q^2} ds \, ,
\label{eq:DI}
\end{equation}
%Eq (11)
where
$N_\text{T}=\sqrt{2}f_\pi/3$, as before, while the spectral
density reads
\begin{equation}
 %\Ds
  N_\text{T}~\rho(Q^2,s)
\equiv
  \frac{1}{\pi} \Im\bigg\{F^{\gamma^*\pi^0}(Q^2,-s- i \epsilon)
=
  N_\text{T}\left[T^{(2)}(Q^2,-s- i \epsilon)+\text{twist-4}\right] \bigg\} \, .
\label{eq:spect-dens}
\end{equation}
% Eq (12)
In the Born approximation, $\1$ is the only term that contributes to
Eq.\ (\ref{eq:T1d}).
It provides the well-known result \cite{Khodjamirian:1997tk}
\begin{equation}
  \rho(Q^2,s)
=
  \left(
                  \frac{\varphi^{(2)}(x)}{Q^2+s}
                  -\frac{\delta_\text{tw-4}^2}{Q^2}\frac{d}{ds}\varphi^{(4)}(x)
            \right)
                   \Bigg|_{x=s/(Q^2+s)}\, ,
\label{eq:Born}
\end{equation}
%Eq (13)
where $\delta_\text{tw-4}^2$ sets the scale for the twist-four
contribution and $\varphi^{(4)}(x)$ is an effective pion DA of
twist four.
The above expression is induced by the particular discontinuity
of the imaginary part of the Born amplitude $T_0(Q^2,-s;y)$, i.e.,
\begin{equation}
 \label{eq:rhoBorn}
  \Ds \frac{1}{\pi}\Im(T_0(Q^2,-s- i \epsilon;y))
\equiv
  \rho^{(0)}(Q^2,s)
=
  \delta(y-x)/(Q^2+s), ~~\text{where}~x=s/(Q^2+s).
\end{equation}
%Eq (14)
In the framework of FOPT it is clear that higher-order corrections to
$\rho$ also contribute owing to the logarithmic factors in the Born
amplitude $T_0(Q^2,q^2,y)$ \cite{Mikhailov:2009kf,Schmedding:1999ap}.
Our goal below will be to obtain the radiative corrections within the
LCSR formalism on the ground of the results in Eqs.\ (\ref{eq:RGsum})
and (\ref{eq:T1dn0}) obtained by RG summation.
As we will see in the next subsection, this procedure will inevitably
lead to an analytic version of QCD perturbation theory
which amends by construction Landau-type singularities.

\subsection{Key element of the radiative corrections in the dispersive representation}
\label{subsec:rad-corr-disp-rel}
The RG summation of all radiative corrections to the TFF in
Eq.\ (\ref{eq:RGsum}) provides another possibility to extract the
imaginary part of the TFF and get the spectral density $\rho$
\cite{Khodjamirian:1997tk,Mikhailov:2009kf,Mikhailov:2016klg}.
Indeed, for the Born contribution, the corresponding imaginary part
is generated by the singularity of $ T_0(Q^2,-s;y)$
(multiplied by powers of logarithmic terms, see Eq.\ (\ref{eq:Born})
and the text below it), while the imaginary part after the RG summation
of the radiative corrections originates also from the
$\Im\left(\bar{a}^\nu_s(-s\bar{y}+Q^2y)/\pi\right)$
contributions.

The general expression for the key perturbative element in this
procedure follows from the first term in Eq.\ (\ref{eq:T1d})
and amounts to the following integral, termed $I_n$:
\begin{subequations}
\begin{eqnarray}
\!\!\!\!&& T_0(Q^2,q^2;y)\left(\bar{a}^{\nu_n}_s(y)\right)
\xlongrightarrow{q^2\to -s}
   \frac{1}\pi \int_{0}^\infty ds \frac{\Im\big[T_0(Q^2,-s;y)\bar{a}^{\nu_n}_s(-s\bar{y}+Q^2y)\big]}{s+q^2}
=
  I_n(Q^2,q^2;y)
\label{eq:1}\\
\!\!\!\!&&=
  \frac{1}{\pi} \int_{0}^\infty \!ds
    \left\{\frac{\Re[T_0(Q^2,-s;y)]\Im[\bar{a}^{\nu_n}_s(-s\bar{y}+Q^2y)]}{s+q^2}
          +\!\frac{\Im[T_0(Q^2,-s;y)]\Re[\bar{a}^{\nu_n}_s(-s\bar{y}+Q^2y)]}{s+q^2}
    \right\}\,.
\label{eq:2}
\end{eqnarray}
%Eq (15a), (15b)
The contribution to the partial TFF $F_n(Q^2,q^2)$ can be expressed
in the form of a convolution between the term $I_n \otimes \Phi_n$,
contained in $I_n(Q^2,q^2;y)$, and the remainder of the partial
harmonic $\Phi_n(y)$.
The latter includes either the next radiative correction
$\Phi_n(y)= \mathcal{T}^{(1)}(y,x)\otimes \psi_n(x)$, or the Born term
$\Phi_n(y)= \1 \otimes \psi_n(x)$---see Eq.\ (\ref{eq:T1d}).
After changing the integration variable
$s \to\sigma = -(-s\bar{y}+Q^2y) \geq 0$,
and applying the principal-value prescription
\begin{equation}
 1/(\sigma +i \varepsilon)= \text{p.v.}\left(1/\sigma\right) - i\pi \delta(\sigma)\, ,
\nonumber
\end{equation}
we set
$T_0(Q^2,-s;y)\sim  -1/(\sigma +i \varepsilon)$
to obtain for the integral in Eq.\ (\ref{eq:2}) the expression
\begin{eqnarray}
 \label{eq:2a}
 I_n(Q^2,q^2;y)
&=&
  -\int_{0}^\infty \frac{d\sigma}{(\sigma+Q(y))}\left\{
  \text{p.v.}\left(\frac{1}{\sigma}\right)
  \theta(\sigma)~\frac{\Im[\bar{a}^{\nu_n}_s(-\sigma)]}{\pi}
  -\delta(\sigma)~\Re[\bar{a}^{\nu_n}_s(-\sigma)]
  \right\}\, .
\end{eqnarray}
\end{subequations}
%Eq (15c)
The second term in Eq.\ (\ref{eq:2a}) containing the
$\delta(\sigma)$ function is induced by the singularity of
$\Im[T_0(Q^2,-s;y)]$ in Eq.\ (\ref{eq:2}).
But, in contrast to the Born case in Eq.\ (\ref{eq:rhoBorn}), this
contribution vanishes for the running coupling
$\Re[\bar{a}^{\nu_n}_s(-0)]=0 $.
This can be explicitly seen in the case of the one- and two-loop
running.
Therefore, only the first term in Eq.\ (\ref{eq:2a}) survives, where
$
 \Ds \frac{1}{\pi}\Im[\bar{a}^\nu_s(-\sigma-i\varepsilon)]
=
 \rho_\nu(\sigma)
$
defines the FAPT spectral density $\rho_\nu(\sigma)$, i.e.,
\begin{eqnarray}
 I_n(Q^2,q^2;y)
&=&
  -\int_{0}^\infty d\sigma \frac{\rho_{\nu_n}(\sigma)}{(\sigma+Q(y))\sigma}\, .
\label{eq:3}
\end{eqnarray}
%Eq (16)
Thus, the key element $I_n(Q^2,q^2;y)$ can be expressed in terms of the
corresponding FAPT couplings.
For the perturbative spectral density $\rho_{\nu}$, we employ the
standard FAPT expression \cite{Bakulev:2005gw}
\begin{eqnarray}
\label{eq:rho-fapt}
  \rho_{\nu}^{(l)}(\sigma)
=
  \frac{1}{\pi}\,
  \textbf{Im}\,\big[a^{\nu}_{(l)}(-\sigma)\big]
=
  \frac{1}{\pi}\,\frac{\sin[\nu~
  \varphi_{(l)}(\sigma)]}{\left(R_{(l)}(\sigma)\right)^{\nu}}~
\xlongrightarrow{\text{1-loop}} \rho_{\nu}^{}(\sigma)=
  \frac{1}{\pi}\,
  \frac{\sin\left[\nu~\arccos\left(L_{\sigma}/\sqrt{L^2_\sigma+\pi^2}\right)\right]}%
  {\beta_0^\nu~\left[\pi^2+L^2_\sigma\right]^{\nu/2}}
 \, ,
\label{eq:spec-dens-nu-C2}
\end{eqnarray}
%Eq (17)
where the phase $\varphi_{(l)}$ and the radial part $R_{(l)}$ have an
$l$-loop content (see \cite{Bakulev:2006ex} and Appendix \ref{App:B}),
whereas
$L_\sigma=\ln(\sigma/\Lambda^2_\text{QCD})$.

Having in mind further considerations to be exposed later, we define
here a more general class of spectral densities $\rho_\nu(m^2;\sigma)$
by inventing a possible gap in the variable $\sigma$, expressed via
the scale $m^2 \geqslant 0$,
$
 \rho_\nu(\sigma) \to \rho_\nu(m^2;\sigma)
=
 \theta(\sigma>m^2)\rho_\nu(\sigma)
$.
Inserting $\rho_\nu(m^2;\sigma)$ into Eq.\ (\ref{eq:3}), we obtain for
$ I_n(Q^2,q^2;m^2)$ the dispersion integral at the lower limit $m^2$,
notably,
\begin{eqnarray}
  I_n(Q^2,q^2,m^2;y)=
- \int_{m^2}^\infty d\sigma \frac{\rho_\nu(\sigma)}{(\sigma+Q(y))\sigma}
&=&
 T_0(Q^2,q^2;y)\Big[\I_{\nu}(m^2,Q(y)) - \M_\nu(m^2)\Big]\,.
\label{eq:4a}
\end{eqnarray}
%Eq (18)
Now the RHS of $I_n(Q^2,q^2,m^2;y)$ in Eq.\ (\ref{eq:4a})
can be decomposed in terms of a new coupling $\I_{\nu}$ and the standard
FAPT couplings $\M_{\nu}$ and $\A_{\nu}$, known from
\cite{Bakulev:2005gw,Bakulev:2006ex}:
\begin{subequations}
\label{eq:4}
\begin{eqnarray}
&& - \int_{Y}^\infty ds \frac{\rho_\nu(s)}{s(s+X)}
=
 \frac{1}{X}\left[\int_{Y}^\infty ds \frac{\rho_\nu(s)}{s+X} -
 \int_{Y}^\infty ds \frac{\rho_\nu(s)}{s}\right]
=
  \frac{1}{X}\left[\I_{\nu}(Y,X) - \M_\nu(Y)\right]
\label{eq:4b}\\
&&
  \I_{\nu}(Y,X)
\stackrel{def}{=}
  \int_{Y}^\infty \frac{d\sigma}{\sigma+X} \rho_{\nu}^{(l)}(\sigma)
\label{eq:4c}\\
&&
  \A_{\nu}(X)
=
  \I_{\nu}(Y \to 0,X),~ \M_{\nu}(Y)
=
  \I_{\nu}(Y,X \to 0),
~\A_{1}(0)
=
  \M_{1}(0)=\I_{1}(Y \to 0,X \to 0)\, .
\label{eq:4d}
\end{eqnarray}
\end{subequations}
%Eq (19a), (19b), (19c)
Note that the structure of a subtraction in the square brackets in
Eq.\ (\ref{eq:4a}) follows from the decomposition of the integrand
in the RHS of (\ref{eq:4b}).
Some remarks regarding Eq.\ (\ref{eq:4}) are here in order.
First, the FAPT couplings $\mathcal{A}_\nu$ \cite{Bakulev:2005gw}
and $\mathfrak{A}_\nu$ \cite{Bakulev:2006ex} refer to the spacelike
and the timelike regime, respectively.
Second, the integral $\I_{\nu}(y,x)$ represents a generalization of the
previous two FAPT couplings, as it becomes obvious from
Eq.\ (\ref{eq:4d}) and from the detailed exposition in
Appendix \ref{App:B}.

We close this discussion by formally defining a new effective coupling
$\bm{\mathbb{A}_\nu}$,
already encountered in (\ref{eq:4a}),
that will be used in the next sections, viz.,
\begin{subequations}
\label{eq:5}
\begin{eqnarray}
\label{eq:eff-coupl}
  \mathbb{A}_\nu(m^2,y)
=
 \I_{\nu}(m^2,Q(y))-\M_\nu(m^2)\,.
\end{eqnarray}
%Eq (20a)
The derivation of $\mathbb{A}_\nu(m^2,y)$ in terms of the new
FAPT coupling $\I_{\nu}$ represents a novelty of the present approach.
At the same time $\mathbb{A}_\nu$ bears through it a process dependence
stemming from the Born term.
This dependence enters via the arguments $Q(y)$, like in the original
case of $\bar{a}_s(y)$, but also through the argument $m^2$ at the
lower limit of the dispersion integral.
Note that for $m^2 \to 0$, one has
\begin{eqnarray}
\mathbb{A}_\nu(0,y) = \A_{\nu}(Q(y)) - \A_\nu( 0)
\end{eqnarray}
\end{subequations}
%Eq (20b)
due to Eq.\ (\ref{eq:4d}).

\subsection{Pion-photon TFF within FAPT}
\label{subsec:TFF-FAPT}
Before we continue let us
i) summarize our findings for the dispersion integral $I_n$,
ii) construct a particular version of the TFF,
and iii) consider it at different scales.

i)
The general expression for
$F^{\gamma^\ast\pi}_\text{FAPT}(Q^2,q^2;m^2)$,
that includes all involved scales, reads
\begin{subequations}
\label{eq:GTFF}
\begin{eqnarray}
\label{G0TFF}
  \nu(n=0)=0;  ~~ F^{\gamma^\ast\pi}_\text{FAPT,0}(Q^2,q^2;m^2)
&=\!&\!
  N_\text{T} T_{0}(Q^2,q^2;y)
\underset{y}{\otimes}
  \left\{\1 +\mathbb{A}_{1}(m^2,y) \mathcal{T}^{(1)}(y,x) \right\}
\underset{x}{\otimes}\psi_0(x) \\
\label{GnTFF} \!
  \nu(n\neq0)\neq0;  ~~ F^{\gamma^\ast\pi}_\text{FAPT,n}(Q^2,q^2;m^2)
&=\!&\!
  \frac{N_\text{T}}{a_s^{\nu_n}(\mu^2)}
  T_0(Q ^2,q^2;y)
\underset{y}{\otimes} \nonumber \\
&&\phantom{\frac{N_\text{T}}{a_s^{\nu_n}(\mu^2)}}
  \left\{\mathbb{A}_{\nu_n}(m^2,y)\1
  +\mathbb{A}_{1+{\nu_n}}(m^2,y) \mathcal{T}^{(1)}(y,x) \right\}
  \underset{x}{\otimes}\psi_n(x)\, .
\end{eqnarray}
\end{subequations}
%Eq (21a), (21b)
It is instructive to compare the above results with the initial
expressions given by Eqs.\ (\ref{eq:T1d}) and (\ref{eq:T1dn0}).
One observes that Eqs.\ (\ref{G0TFF}) and (\ref{GnTFF}) have the
same structure as the original expressions and can be recast
into the form of Eq.\ (\ref{eq:T1d}) and Eq.\ (\ref{eq:T1dn0}),
respectively, using the evident replacement
$\mathbb{A}_{\nu}(m^2,y) \to \bar{a}_s^{\nu}(y)$.

ii)
We show next the results for
$F^{\gamma\pi}_\text{FAPT}(Q^2;m^2)$ in
the limits
$q^2 \to 0$, $Q(y) \to yQ^2$ and $m^2 \geqslant 0$ in explicit form:
\begin{subequations}
\label{eq:TFFq0}
\begin{eqnarray}
\label{n0pTFFq0}
  \nu(n=0)=0; && Q^2 F^{\gamma\pi}_{\text{FAPT},0}
\equiv
  F_0(Q^2;m^2)
=
  N_\text{T}\left\{\int^1_{0} \!\! \frac{\psi_0(x)}{x}~dx +
  \left(\frac{\mathbb{A}_{1}(m^2,y)}{y}\right)
\underset{y}{\otimes}
  \mathcal{T}^{(1)}(y,x)
\underset{x}{\otimes}
  \psi_0(x)\right\} \, ,\\
\nu(n\neq0)\neq0; &&  Q^2 F^{\gamma\pi}_{\text{FAPT},n}
\equiv
  F_n(Q^2;m^2)
=
\nonumber \\
&&  \frac{N_\text{T}}{a_s^{\nu_n}(\mu^2)}
  \left\{\left(\frac{\mathbb{A}_{\nu_n}(m^2,y)}{y}\right)
\underset{y}{\otimes}
  \psi_n(y)+\left(\frac{\mathbb{A}_{1+{\nu_n}}(m^2,y)}{y}\right)
\underset{y}{\otimes}
  \mathcal{T}^{(1)}(y,x)
\underset{x}{\otimes}
  \psi_n(x)\right\} \, .
\label{npTFFq0}
\end{eqnarray}
\end{subequations}
%Eq (22a), (22b)
These equations can again be related to the initial expressions
given by Eqs.\ (\ref{eq:T1d}) and (\ref{eq:T1dn0}) by means of
the replacement
$
 \mathbb{A}_{\nu}(m^2,y)
=
 \I_{\nu}(m^2, Q^2y)-\M_\nu(m^2) \rightarrow \bar{a}_s^\nu(y)
$.

iii)
Definition (\ref{eq:eff-coupl}) of the effective coupling
$\mathbb{A}_{\nu}$, supplemented by
Eqs.\ (\ref{eq:GTFF}) and (\ref{eq:TFFq0}),
reveals that the high-energy asymptotic behavior of the form-factor
components $F_n(Q^2)$ is determined in part by the low-energy behavior
of $\M_{\nu}(m^2)$ or the value of $\M_{\nu}(0)=\A_{\nu}(0)$
for $m^2=0$.
Thus, determining the \textit{low-energy behavior} of the FAPT
couplings, one would be in the position to extract information about
the \textit{high-energy behavior} of the transition form factor---an
arguably unexpected result that demands a rigorous explanation.

\subsection{Low-energy modification of FAPT---calibration procedure}
\label{subsec:mod-FAPT}
\begin{figure}[hbt] %\unitlength=1mm
%\centering\epsfig{file=mb(mu).eps,width=8.cm}
\includegraphics[width=.45\linewidth]{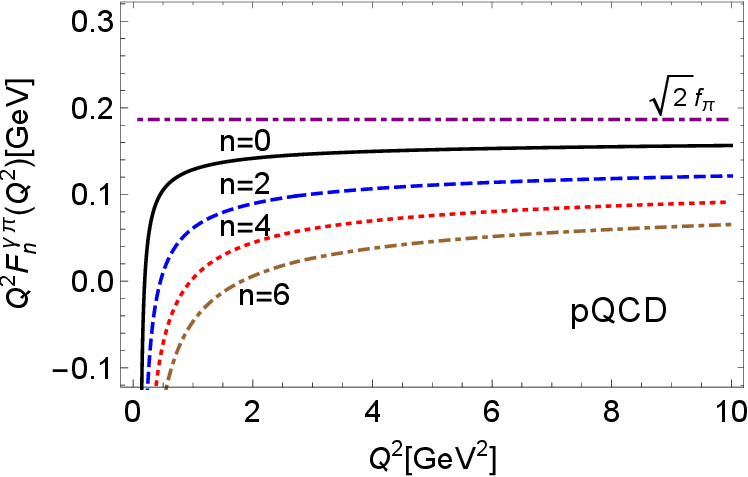}
~~~~~\includegraphics[width=.45\linewidth]{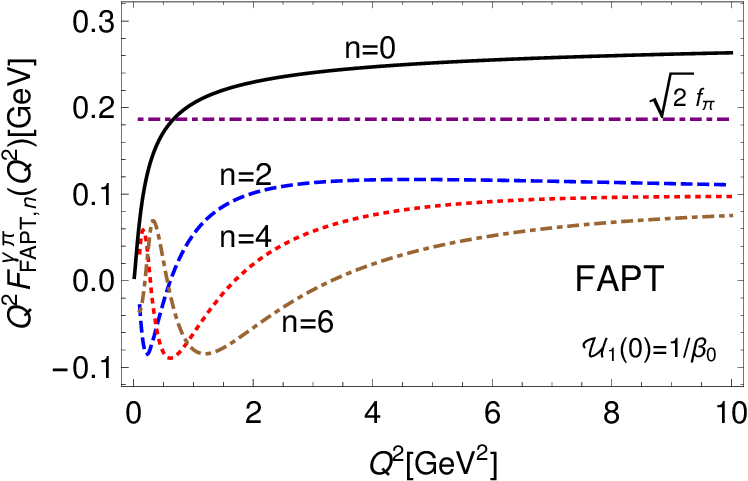}
%\vspace{-0.5cm}
\caption{\textit{Left panel}: $Q^2 F^{\gamma\pi}_{n}(Q^2)$ for $n=0,2,4,6$
                        calculated in pQCD;
                        \textit{Right panel}: The same quantity calculated with the modified
                        (but uncalibrated)
                        FAPT, $Q^2 F^{\gamma\pi}_{\text{FAPT},n}(Q^2)$, according to
                        Eq.\ (\ref{eq:TFFq0}) for
                        ${\mathfrak A}^{(1)}_{1}(0)=1/\beta_0$
                        with $m^2 =4 m_\pi^2 \approx 0.08$ GeV$^2$.
\label{fig:fig1}}
\end{figure}

However, this seemingly obvious connection doesn't work
because it causes a spurious contribution to the
asymptotic value of the TFF that contradicts pQCD.
Indeed, for the effective couplings of analytic perturbation
theory (APT) one has according to (\ref{eq:A-def})
$\mathbb{A}_{1}=\I_{1}(m^2, Q^2y)-\M_1(m^2) < 0$, which
entails a sign flip of the radiative corrections due to the
second term $- \M_1(m^2)$.
Moreover, for $m^2=0$,
${\cal A }^{(1)}_{1}(0)={\mathfrak A}^{(1)}_{1}(0)=1/\beta_0$
\cite{Shirkov:1997wi,Milton:1996fc,Milton:1997us}
and $\mathbb{A}_1(0,y) \to \left[\A_{1}(Q(y))-\A_1( 0)\right] $.
If one would substitute these expressions into Eq.\ (\ref{n0pTFFq0}),
one would immediately arrive at a result for the asymptotic
(scaled) TFF that would clearly contradict the limit derived with
pQCD in the asymptotic regime.
This contradiction in the behavior of $Q^2F_0(Q^2\to \infty)$ can
be traced back to the radiative corrections to the inverse moment
that are created just by the disturbing term $-\A_1( 0)$.
In fact, the deviation from the standard contribution,
proportional to $a_s(\mu^2 \sim Q^2)$, can be estimated from
Eq.\ (\ref{n0pTFFq0}), see \cite{Mikhailov:2009kf}, to be
given by the following distortion term
\begin{eqnarray}
\!\!\!\delta=-\left(\frac{\A_1( 0)}{y}\right)
\underset{y}{\otimes}
  \mathcal{T}^{(1)}(y,x)
\underset{x}{\otimes}
  \psi_0(x)
=
  -\A_1(0) C_\text{F}\int_0^1 \frac{dx}{x}
  \left[-5 +\frac{\pi^2}{3}- \ln^2\left(\bar{x}/x\right) \right]\psi_0(x)
=
  15\A_1(0)C_\text{F}\, .
\end{eqnarray}
%Eq (23)
The quantity $\delta$ is a large constant, comparable to the Born term
$
 \Ds \int^1_{0} \!\! \frac{\psi_0(x)}{x}~dx
=
 3\sim 15 \A_1(0) C_\text{F}
=
 \frac{1}{\beta_0}20=$
$\Ds \frac{20}{9}$
(with $N_f=3$),
that certainly destroys the asymptotic behavior of $Q^2F_{n=0}(Q^2)$.
The latter turns out to be larger than the asymptotic limit
$\sqrt{2}f_\pi\approx 0.187$
already in the vicinity of the normalization scale $\mu_0\simeq 1$~GeV,
as one can see from the behavior of the solid line in the right panel
of Fig.\ \ref{fig:fig1}, and disagrees with the analogous result
obtained in the left panel within pQCD.
Such a situation calls for a remedy.

To restore the correspondence of our perturbative expansion to the
standard pQCD theory, we have to eliminate the distortion term to
the TFF at $Q^2\to\infty$ by appropriately adjusting the mathematical
IR behavior of the new analytic couplings in such a way as to preserve
the validity of the QCD asymptotics applicable to this particular
physical process.
This can be achieved by imposing the condition
$\A_{1}(0),~\M_{1}(0) \simeq 0$ in $\mathbb{A}_{1}$,
which is tantamount to a ``calibration'' of their behavior
in Eq.\ (\ref{eq:eff-coupl}).
%%%%%%%%%%%%%%%%%%%%%%%%%%%%%%%%%%%%%%%%%%%%%%%%%%%%%%%%%%%%%%%%%%%%%%%%%%%%%%%%%%%%%%%%%%%%%%%%%%% Figure 1
%%%%%%%%%%%%%%%%%%%%%%%%%%%%%%%%%%%%%%%%%%%%%%%%%%%%%%%%%%%%%%%%%%%%%%%%%%%%%%%%%%%%%%%%%%%%%%%%%%%

\begin{figure}[htb] %\unitlength=1mm
%\centering\epsfig{file=mb(mu).eps,width=8.cm}
\includegraphics[width=0.45\linewidth]{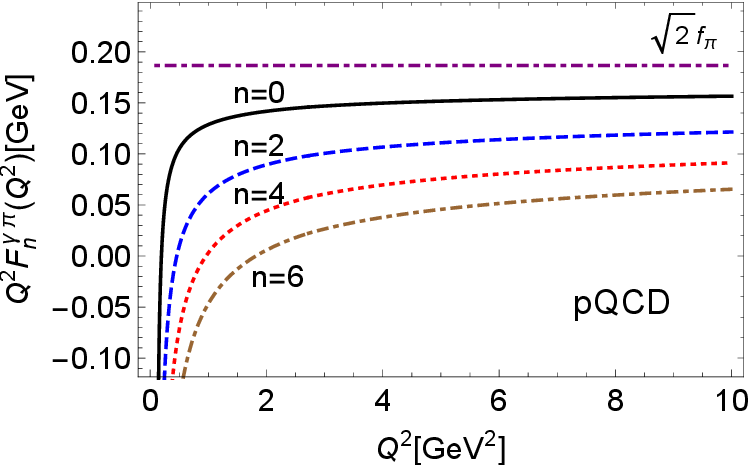}
~~~~~\includegraphics[width=0.45\linewidth]{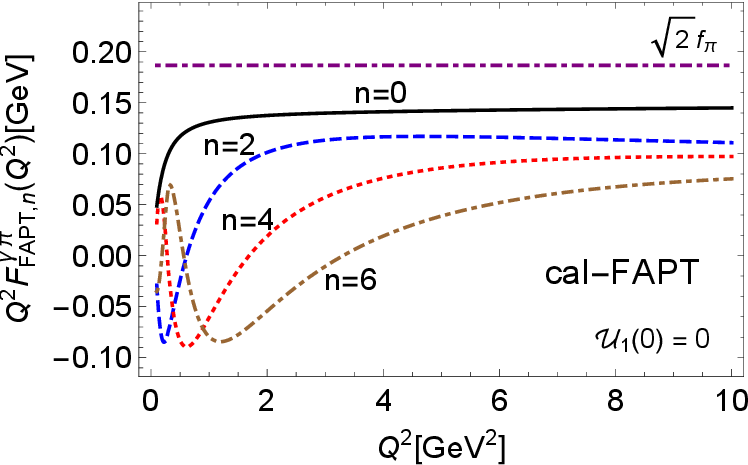}
\vspace{0.1cm}
\caption{\textit{Left panel}: $Q^2 F^{\gamma\pi}_n(Q^2)$ for $n=0,2,4,6$
                        calculated in pQCD as in Fig.\ \ref{fig:fig1}.
                        \textit{Right panel}: TFF $Q^2 F^{\gamma\pi}_{\text{FAPT},n}(Q^2)$
                        according to Eq.\ (\ref{eq:TFFq0})
                        within the calibrated FAPT $\M_{\nu}(0)=\A_{\nu}(0)=0$.
\label{fig:fig2}
}
\end{figure}

The calibration procedure has more consequences.
Consider noninteger values of the index $\nu$ within FAPT.
Then the \textit{unbounded} behavior of the FAPT couplings at one
loop for $0< \nu <1$,
$
 {\cal A }^{(1-{\rm loop})}_{\nu < 1}(Q^2),
 ~{\mathfrak A}^{(1-{\rm loop})}_{\nu < 1}(Q^2)
$
near $Q^2=0$ would render the corresponding expressions in
Eqs.\ (\ref{GnTFF}), (\ref{npTFFq0}) meaningless
(with the related details being given in Appendix \ref{App:B}).
Consequently, in order to obtain a TFF with the correct
behavior in the asymptotic regime, we have to set
\begin{equation}
\label{eq:zerocondition}
~~{\cal A }^{(1)}_{\nu}(0)
={\mathfrak A}^{(1)}_{\nu}(0)=0,~\text{for}~  0< \nu \leqslant 1
\end{equation}
%Eq (24)
also for $Q^2=0$ for $0< \nu \leqslant 1$.
This implies that the initial FAPT couplings, which were constructed
without employing any nonperturbative input at low momenta to saturate
them in the deep IR regime, should be corrected at
$Q^2 \approx 0$ \textit{a posteriori}
in order to avoid spurious, i.e., unphysical constants.
Thus, the present scheme represents an advance over the original
FAPT and corrects the corresponding expressions for the analytic
couplings displayed in the middle columns of
Eqs.\ (\ref{eq:B6}) in Appendix \ref{App:B}.
Note that the imposition of calibration is a novelty
of the present investigation and differs from other approaches
that try to model the low-energy behavior of the couplings
\cite{Ayala:2012xf,Ayala:2016zrz,Ayala:2017tco}.
The difference relative to \cite{Ayala:2012xf} arises from the
imposition $\A_\nu(0)=\delta$ at small fixed value $\delta$, while
in \cite{Ayala:2016zrz,Ayala:2017tco} one sets $\A_\nu(0)\sim Q^2$
(when $Q^2\mapsto0$) as suggested
by lattice simulations \cite{Bogolubsky:2009dc} for dressing functions.
In any case, the above redefinition based on calibration
renders the TFF compatible with the QCD asymptotic limit for $\nu > 1$,
as one can see from the last column in Eq.\ (\ref{eq:B6}).
In what follows, we will refer to this \textit{calibrated}
version of FAPT as ``cal-FAPT''.
As one sees from the right panel of Fig.\ \ref{fig:fig2}, this version
of FAPT guarantees that the TFF behavior subject to condition
(\ref{eq:zerocondition}) indeed reproduces the pQCD limit and the TFF
result for $n=0$ --- left panel --- in contrast to the $Q^2F_{n=0}(Q^2)$
behavior entailed by the standard FAPT couplings, see Fig.\ \ref{fig:fig1}
(solid line for $n=0$ in the right panel).
In contrast, the calibrated counterpart of the TFF shows excellent
agreement with the PQCD result as it is effected in the right panel of
Fig.\ \ref{fig:fig2}.

%%%%%%%%%%%%%%%%%%%%%%%%%%%%%%%%%%%%%%%%%%%%%%%%%%%%%%%%%%%%%%%%%%%%%%%%%%%%%%%%%%%%%%%%%%%%%%%%%%% Figure 2
%%%%%%%%%%%%%%%%%%%%%%%%%%%%%%%%%%%%%%%%%%%%%%%%%%%%%%%%%%%%%%%%%%%%%%%%%%%%%%%%%%%%%%%%%%%%%%%%%%%%

\section{Hadronic photon content in LCSR}
\label{sec:mesonic-FAPT}
In the previous section we have constructed  a new perturbative expansion
that includes all radiative corrections to the TFF using RG summation while
preserving its QCD asymptotics.
In this section, we are going to implement this scheme to the LCSR formulation
in terms of the calibrated FAPT.

To this end, let us extend the initial expression for the pion TFF in the LCSR,
given by Eq.\ (\ref{eq:DI}), to the case of a quasi-real
photon $\gamma(q^2)$ with a virtuality $q^2 \ll m^2_\rho$.
This can be done in terms of a physical spectral density that takes
into account the vector-meson properties of the quasi-real photon.
Employing the physical spectral density
$\rho^\text{ph}$ with
$
 \rho^\text{ph}(\sigma)
=
 \delta(\sigma-m^2_\rho) \sqrt{2}f_\rho F^{\rho \pi}(Q^2)
 +\theta(\sigma > s_0) N_{\text{T}}\rho(Q^2,\sigma)
$
we substitute $\rho^\text{ph} \to \rho$ into
(\ref{eq:DI}), to obtain \cite{Khodjamirian:1997tk}
\begin{eqnarray}
\label{eq:LCSR}
  F^{\gamma^*\pi}\left(Q^2,q^2\right)
= N_{\text{T}} \int_{s_0}^{\infty} \frac{\rho(Q^2,\sigma) d\sigma}{\sigma+q^2}
  + \sqrt{2}f_\rho \frac{F^{\rho \pi}(Q^2)}{m^2_\rho+q^2}\, ,
\end{eqnarray}
%Eq (25)
where the term
$\delta(s-m^2_\rho)\sqrt{2}f_\rho F^{\rho \pi}(Q^2)$
in the physical spectral density models the $\rho/\omega$-resonances.
Applying  the ``duality approximation'', that involves the TFF
$F^{\rho \pi}$ to describe the intermediate subprocess
$\gamma^*\rho^0 \to \pi^0$
(see Sec.\ 2 in \cite{Khodjamirian:1997tk} for details),
we find
\begin{subequations}
\label{dual}
\begin{eqnarray}
\label{eq:dual}
  \sqrt{2}f_\rho\frac{F^{\rho \pi}(Q^2)}{m^2_\rho+q^2}
=
  N_{\text{T}}
  \int_{0}^{s_0} \frac{ \rho(Q^2,s) ds}{s+q^2} \, .
\end{eqnarray}
%Eq (26a)
From the Borel transform $\hat{B}_{q^2 \to M^2}$ of (\ref{eq:dual}),
we get
\begin{equation}
\label{eq:rhopi}
  \sqrt{2}f_\rho F^{\rho \pi}(Q^2)= N_{\text{T}} \int_{0}^{s_0}
        \exp\left(
                  \frac{m_{\rho}^2-s}{M^2}
            \right)
        \! \rho(Q^2,s)
  ds \, ,
\end{equation}
%Eq (26b)
\end{subequations}
finally arriving at the total TFF expression for
$F^{\gamma^*\pi}$ \cite{Khodjamirian:1997tk},
evaluated in the limit $q^2 \to 0$, i.e.,
\begin{eqnarray}
   F^{\gamma \pi}\left(Q^2\right)
=
  N_{\text{T}}
  \left[ \int_{s_0}^{\infty} \rho(Q^2,s)
        \frac{ds}{s}
 +
        \frac{1}{m_{\rho}^2}
        \int_{0}^{s_0}
        \exp\left(
                  \frac{m_{\rho}^2-s}{M^2}
            \right)
        \! \rho(Q^2,s) ds \right]\, ,
\label{eq:ini-LCSR}
\end{eqnarray}
%Eq (27)
where $M^2$ is the Borel parameter.
We use for simplicity, the fixed value
$M^2\approx0.9$ GeV$^2$.
Increasing the Borel mass to $M^2=1.1$~GeV$^2$ would affect the TFF
between 10 and 40~GeV$^2$ only by about $4\%$ \cite{Stefanis:2012yw},
which shows that its influence on the TFF is small, see also \cite{Agaev:2010aq}.
For a more sophisticated treatment, we refer to our previous works,
e.g., \cite{Bakulev:2011rp}.

The first term in Eqs.\ (\ref{eq:LCSR}) and
analogously in (\ref{eq:ini-LCSR}) stems from the
\textit{hard (i.e., quark-gluon) part}
with the integration taken over the duality interval $s_0$.
Here $s_0$ plays the role of the main scale parameter in
the model of the physical density.
Note that for $s_0 \to 0$, expression (\ref{eq:ini-LCSR}) reduces to
the first term, i.e., to the initial form of Eq.\ (\ref{eq:DI}),
and further to the harmonic expansions encountered in
Eqs.\ (\ref{eq:GTFF}) and (\ref{eq:TFFq0}).
The numerical value of the effective threshold parameter $s_0$ in the
$\rho$-meson channel is fixed to the value $s_0=1.5$~GeV$^2$, see,
for instance, \cite{Bakulev:2002uc}.
The second term in Eq.\ (\ref{eq:ini-LCSR}) is the result obtained for
the TFF $F^{\rho \pi}$ following from Eq.\ (\ref{eq:rhopi})
and originates from the \textit{soft (i.e., hadronic) part} of the
pion TFF.
For the considerations to follow, it is useful to reduce
Eq.\ (\ref{eq:ini-LCSR}) to the Born approximation by taking for
$\rho$ the expression in (\ref{eq:Born}) and by replacing the variable
of integration $\Ds s \to x=s/(Q^2+s)$ to get
\begin{subequations}
\label{eq:BornLCSR}
\begin{eqnarray}
  Q^2 F^{\gamma \pi}\left(Q^2\right)
&=&
  N_{\text{T}}
  \left[\int_{x_0}^{1} \bar{\rho}(Q^2,\bar{x})
        \frac{dx}{x}
 + \!\frac{Q^2}{m_{\rho}^2}
        \int_{0}^{x_{0}}
        \exp\left(
                  \frac{m_{\rho}^2-Q^2x/\bar{x}}{M^2}
            \right)
        \! \bar{\rho}(Q^2,\bar{x})
  \frac{dx}{\bar{x}}\right]\,,\label{eq:LCSR-FQq}\\
&&\bar{\rho}(Q^2,x)=\varphi^{(2)}_{\pi}(x)
  +\frac{\delta_\text{tw-4}^2}{Q^2}x\frac{d}{dx}\varphi^{(4)}(x)\,,
\end{eqnarray}
%Eqs (28a), (28b)
where $x_0=s_0/(Q^2+s_0)$.
The twist-four term depends on the parameter
$\delta_\text{tw-4}^2(\mu^2)$, which assumes values in the interval
$\delta_\text{tw-4}^2(\mu^2)=0.19\pm 0.04$~GeV$^2$ \cite{Bakulev:2002hk}.
Below we shall use the elements of the expansion $\bar{\rho}(Q^2,x)$
expressed in terms of the Gegenbauer harmonics, i.e.,
$
 \bar{\rho}(Q^2,x)
=
 \bar{\rho}_0(Q^2,x)+ \sum_{n=2,4,...} a_n(Q^2)\bar{\rho}_n(Q^2,x)$,
where
\begin{eqnarray}
  \bar{\rho}_0(Q^2,x)
=
  \psi_0(x)+\frac{\delta_\text{tw-4}^2}{Q^2}x\frac{d}{dx}\varphi^{(4)}(x);~\varphi^{(4)}(x)=\frac{80}{3}x^2(1-x)^2;~
  \bar{\rho}_n(Q^2,x)=\psi_n(x)\, .
\end{eqnarray}
\end{subequations}
%Eq (28c)
Note that in this expression we combined the twist-four contribution with the
$\psi_0$ component of the twist-two spectral density into a single
spectral density termed $\bar{\rho}_0$.

\subsection{Hard part of LCSR with RG summation}
\label{subsec:h-LCSR}
The first term in Eq.\ (\ref{eq:ini-LCSR}), notably,
\begin{eqnarray}
\label{eq:H-fapt}
  N_{\text{T}} \int_{s_0}^{\infty}
        \frac{ds}{s}\rho(Q^2,s)
=
  N_{\text{T}}~ F^\text{H}_\text{FAPT}(Q^2;m^2,s_0)\, ,
\end{eqnarray}
%Eq (29)
represents the hard (label H) contribution to the LCSR
and can be directly expressed in terms of FAPT.
The only difference with respect to Eqs.\ (\ref{eq:GTFF}),
(\ref{eq:TFFq0}), is the lower limit of integration $s_0$
instead of zero.
This shift induces a more complicated structure of the effective
coupling $\mathbb{A}_\nu \to \mathbb{A}_\nu(m^2,s_0;y)$ as it now
depends on two scale-parameters: $m^2$ and $s_0$.
Performing similar calculations as those to derive Eq.\ (\ref{eq:4a})
for the key element $I_n$ in Subsection \ref{subsec:rad-corr-disp-rel},
we derive the following expression
\begin{subequations}
\label{eq:eff-coupl20}
\begin{eqnarray}
  \mathbb{A}_\nu(m^2,s_0;y)
\!\!&=&\!\!
  \Ds \theta\left(y\geqslant y_0\right)
  \left[\I_{\nu}(m^2,Q(y))-\M_\nu(m^2)\right]
  + \theta\left(y < y_0\right)\left[\I_{\nu}(s_0(y),Q(y))-\M_\nu(s_0(y))\right],
\label{eq:eff-coupl2} \\
&&s_0(y)=
  s_0 \bar{y}-Q^2y,~~~
  y_0= \frac{s_0-m^2}{s_0+Q^2}\, ,
\label{eq:eff-coupl2-cond}
\end{eqnarray}
%Eq (30a), (30b)
where we have imposed the condition $m^2<s_0$ (see for the
general expression given by Eq.\ (\ref{eq:eff-coupl2-C5}) Appendix \ref{App:C}).
The effective coupling $\mathbb{A}_\nu(m^2,s_0;y)$ is a continuous
function in the vicinity of $y=y_0$ and $s_0(y_0)=m^2$, as it
follows from the definitions in Eq.\ (\ref{eq:eff-coupl2-cond}).
Note that in the limit $m^2=0$,
$\mathbb{A}_\nu(0,s_0;x)$ in Eq.\ (\ref{eq:eff-coupl2}) becomes
\begin{eqnarray}
 \mathbb{A}_\nu(0,s_0;x)
\!\!&=&\!\!
  \Ds \theta\left(x\geqslant x_0\right)
  \left[\A_{\nu}(Q(x))-\A_\nu(0)\right]
  + \theta\left(x < x_0\right)\left[\I_{\nu}(s_0(x),Q(x))-\M_\nu(s_0(x))\right]\,,
\label{eq:eff-coupl3}
\end{eqnarray}
\end{subequations}
%Eq (30c)
where $\M_\nu(0)=\A_\nu(0)$ and $y_0=x_0=s_0/(Q^2+s_0)$.
Let us conclude these considerations by presenting the harmonic
representation for the hard part in Eq.\ (\ref{eq:H-fapt}):
\begin{subequations}
\label{eq:TFFq02}
\begin{eqnarray}
\label{n0pTFFq2}
 \!\! Q^2 F^\text{H}_{\text{FAPT},0}(Q^2;m^2,s_0)
&=&
%\!\!&\!\!=\!\!&\!\!
  N_\text{T} \left\{\int^{1}_{x_0} \!\! \bar{\rho}_0(Q^2,\bar{x})\frac{dx}{x} +
  \left(\frac{\mathbb{A}_{1}(m^2,s_0;x)}{x}\right)
\underset{x}{\otimes}
  \mathcal{T}^{(1)}(x,y)
\underset{y}{\otimes}
  \psi_0(y)  \right\}, \\
 \!\!Q^2 F^\text{H}_{\text{FAPT},n}(Q^2;m^2,s_0)
&=&
\frac{N_\text{T}}{a_s^{\nu_n}(\mu^2)}
  \left\{\left(\frac{\mathbb{A}_{\nu_n}(m^2,s_0;x)}{x}\right)
\underset{x}{\otimes}
  \1+\left(\frac{\mathbb{A}_{1+{\nu_n}}(m^2,s_0;x)}{x}\right)
\underset{x}{\otimes}
  \mathcal{T}^{(1)}(x,y)\right\}
\underset{y}{\otimes}
  \psi_n(y)\, .
\label{npTFFq2}
\end{eqnarray}
%Eq (31a), (31b)
\end{subequations}
The first entry in Eq.\ (\ref{n0pTFFq2}) pertains to the
contribution of the zero-harmonic in the expansion of the first
term in Eq.\ (\ref{eq:LCSR-FQq}).
The structure of the next entry in Eq.\ (\ref{n0pTFFq2})---which
represents the radiative correction related to $\mathbb{A}_{1}$---is
more interesting because the first term in Eq.\ (\ref{eq:eff-coupl3})
(simplified by setting $m^2=0$)
corresponds to the integration over the expected hard region
$x \geqslant x_0$,
while in the region $x < x_0$ a new contribution from the
second term appears in addition.
On the other hand, the limit $x_0\to 0$, $s_0\to m^2$ (or 0)
in the hard part of Eqs.\ (\ref{n0pTFFq2}), (\ref{npTFFq2}) reproduces
the known FAPT result encountered in Eqs.\ (\ref{n0pTFFq0}), (\ref{npTFFq0}).

\subsection{Soft part of LCSR with RG summation}
\label{subsec:s-LCSR}
We consider now the soft part of the LCSR expressed by (\ref{eq:ini-LCSR}).
The quasi-real photon induces a contribution that is encoded in
$\sqrt{2}f_\rho F^{\rho \pi}$ and can be expressed within FAPT as
follows
\begin{eqnarray}
  \sqrt{2}f_{\rho}F^{\rho \pi}(Q^2)
&=&
  \Ds N_{\text{T}}\exp\left(\frac{m_{\rho}^2}{M^2}\right)~\hat{B}_{q^2 \to M^2}
  \left[
        \int_{m^2}^{s_0} \frac{\rho(Q^2,s) ds}{s+q^2}
        =
        F^{\gamma^\ast\pi}_\text{FAPT}(Q^2,q^2;m^2)
        - F^{\gamma^\ast\pi}_\text{FAPT}(Q^2,q^2; s_0)
  \right]\, .
\label{eq:res-fapt}
\end{eqnarray}
%Eq (32)
The term
$F^{\gamma^\ast\pi}_\text{FAPT}(Q^2,q^2;m^2)$
was already discussed in connection with Eq.\ (\ref{eq:GTFF}).
Taking it into account and employing the definition of the
effective coupling, we obtain
\begin{subequations}
 \label{eq:32}
\begin{eqnarray}
   &&\exp\left(\frac{m_{\rho}^2}{M^2}\right)
   ~\hat{B}_{q^2 \to M^2}\left[
                               F^{\gamma^\ast\pi}_\text{FAPT}(Q^2,q^2;m^2)
                               - F^{\gamma^\ast\pi}_\text{FAPT}(Q^2,q^2;s_0)
                         \right]
=
\nonumber \\
n=0: &&     \int^{x_0}_{0}
  \exp\left(
            \Ds \frac{m_{\rho}^2}{M^2}- \frac{Q^2}{M^2}\frac{x}{\bar{x}}
      \right)
            \bar{\rho}_0(Q^2,\bar{x})\frac{dx}{\bar{x}}
\label{eq:32a}\\
&& +        \int^{x_0}_{0}\exp\left(
            \Ds \frac{m_{\rho}^2}{M^2}- \frac{Q^2}{M^2}\frac{x}{\bar{x}}
            \right)\frac{dx}{\bar{x}}\Delta_1(m^2,x)
            \mathcal{T}^{(1)}(x,y) \underset{y}{\otimes}\psi_0(y)+ O(\mathbb{A}_2)
\label{eq:21}
\end{eqnarray}
\begin{eqnarray}
n\neq 0:&& \int^{x_0}_{0} \exp\left(
                                   \Ds \frac{m_{\rho}^2}{M^2}- \frac{Q^2}{M^2}\frac{x}{\bar{x}}
                             \right)
                                   \frac{dx}{\bar{x}}
            \bigg[\Delta_{\nu_n}(m^2,x)\psi_n(x) %\\
            +\Delta_{1+\nu_n}(m^2,x)
            \mathcal{T}^{(1)}(x,y) \underset{y}{\otimes}\psi_n(y)\bigg]+ O(\mathbb{A}_2) \, ,
\label{eq:23}
\end{eqnarray}
\end{subequations}
%Eq (33a), (33b), (33c)
where the first term in Eq.\ (\ref{eq:32a}) corresponds to the
zero-harmonics part of the second term in Eq.\ (\ref{eq:LCSR-FQq}).
Note that a new coupling
$\Delta_{\nu}(m^2,y)$
appears in these equations, which originates from the differences
$[\mathbb{A}_{\nu}(m^2;x)-\mathbb{A}_{\nu}(m^2,s_0;x)]$
written in the form
\begin{subequations}
 \label{eq:Aeffect}
 \begin{eqnarray}
  \mathbb{A}_{\nu}(m^2;y)-\mathbb{A}_{\nu}(m^2,s_0;y)
&=&\theta(y < y_0)~\Delta_{\nu}(m^2,y)
\nonumber \\
  \Delta_{\nu}(m^2,y)&=& \left[\I_{\nu}(m^2,Q(y))-\I_{\nu}(s_0(y),Q(y))+\M_\nu(s_0(y))-\M_\nu(m^2)\right] \,,
\label{eq:Aeffect-m} \\
  \mathbb{A}_{\nu}(0;x)-\mathbb{A}_{\nu}(0,s_0;x)
&=& \theta(x < x_0)~\Delta_{\nu}(0,x) \nonumber \\
\Delta_{\nu}(0,x)&=& \left[\A_{\nu}(Q(x))-\I_{\nu}(s_0(x),Q(x))+\M_\nu(s_0(x))-\M_\nu(0)\right]\,
\label{eq:Aeffect-m=0}\, .
\end{eqnarray}
\end{subequations}
%Eqs (34a), (34b)
One can see from Eqs.\ (\ref{eq:32}) and (\ref{eq:Aeffect}) that the
integration domain $x<x_0$ for the radiative corrections is the same as for the
Born term in (\ref{eq:32a}).
Moreover, the standard structure of the integral is restored,
while the quantity $\Delta_{\nu}$ plays the role of an effective coupling
in the soft part.

Expressions (\ref{eq:21}), (\ref{eq:23}) are derived by assuming that the
effective coupling $\mathbb{A}_{\nu}$ does not depend
(by means of the function $Q(y)$) on $q^2$
and is taken at $q^2=0$, i.e., $Q(y) \to yQ^2$.
This approximation is justified, provided $\mathbb{A}_{\nu}(m^2,y)$
depends on $q^2$ in a significantly weaker way than $T(Q^2,q^2;y)$.
We shall consider the conditions for the validity of this approximation
farther below.

Let us now concentrate our attention on the complete LCSR result for
the TFF by combining the soft part (denoted by S) in (\ref{eq:32})
 \begin{eqnarray}
  F^\text{S}_\text{FAPT}\left(Q^2\right)
&=&
        \frac{1}{m_{\rho}^2}
        \exp\left(\frac{m_{\rho}^2}{M^2}\right)
        \hat{B}_{q^2 \to M^2}\left[ F^{\gamma^\ast\pi}_\text{FAPT}(Q^2,q^2;m^2)
        - F^{\gamma^\ast\pi}_\text{FAPT}(Q^2,q^2; s_0) \right]
\label{eq:res-fapt-a}
 \end{eqnarray}
%Eq (35)
with the hard part given by Eqs.\ (\ref{eq:H-fapt}) and
(\ref{eq:TFFq02}).
Substituting Eqs.\ (\ref{eq:H-fapt}) and (\ref{eq:res-fapt-a}) into the
initial Eq.\ (\ref{eq:LCSR}), we arrive at the final equation for
$F^{\gamma^*\pi}_\text{LCSR}\left(Q^2\right)$:
\begin{eqnarray}
   Q^2 F^{\gamma \pi}_\text{LCSR}\left(Q^2\right)
&=&
  N_{\text{T}}
  \bigg[Q^2 F^\text{H}_\text{FAPT}\left(Q^2\right)+
        Q^2 F^\text{S}_\text{FAPT}\left(Q^2\right)
       + \text{twist-4} \bigg] \, .
\label{eq:res-fapt-b}
\end{eqnarray}
%Eq (36)
This equation is the FAPT analogue of the LCSR given
by Eq.\ (\ref{eq:ini-LCSR}).
However, it possesses some advantages relative to that and includes
some new effects:
(i) It has no singularities in the perturbative expansion
(cf. Eq. (\ref{eq:RGsum})).
(ii) It contains all logarithmic power corrections by virtue of the
RG-summation (taken here in the one-loop approximation), while the
values of these corrections are smaller in comparison to those in
standard pQCD using FOPT---thus significantly improving the
reliability of the perturbative expansion.
(iii) It takes into account the particular low-energy behavior
of the quasi-real photon within the LCSR approach.

\section{Effect of RG summation on the TFF in LCSR}
\label{sec:phenomenology}
From the calculational point of view, this FAPT-inspired approach may
help avoid the appearance of large radiative corrections to the
pion-photon TFF at low/moderate momenta because such terms become
small by virtue of the FAPT summation in contrast to FOPT.
Indeed, the smallness of the next-order FAPT coupling $\I_{2}$ in
Fig.\ \ref{fig6}, shown in App.\ \ref{App:B}
by a shaded area at the bottom of the 3D graphics
explicitly illustrates this feature.

The content of the RHS of the general LCSR expression
(\ref{eq:res-fapt-b}) for the pion-photon TFF can be recast in
terms of the $\psi_n$--expansion,
cf.\ Eq.\ (\ref{eq:FFgegen-exp}), to obtain
\begin{eqnarray}
\label{eq:37}
 \!\! \!\! F^{\gamma\pi}_\text{LCSR}\left(Q^2\right)
=
 F^{\gamma\pi}_{\text{LCSR};0}\left(Q^2\right)
 + \sum_{n=2,4,\ldots} a_n(\mu^2)~F^{\gamma\pi}_{\text{LCSR};n}\left(Q^2\right),
\end{eqnarray}
%Eq (37)
where the expressions for the partial TFFs
$F^{\gamma\pi}_{\text{LCSR};n}$
follow from Eqs.\ (\ref{eq:TFFq02}) and (\ref{eq:32}).
To simplify the final representation of
$F^{\gamma \pi}_{\text{LCSR};n}$,
as well as for further analysis in future work,
we use the effective couplings
$\mathbb{A}_{\nu}(0,s_0,y)$ for $m^2=0$
from Eq.\ (\ref{eq:eff-coupl3}) and Eq.\ (\ref{eq:Aeffect-m=0})
so that we arrive at
\begin{subequations}
 \label{eq:finalFAPTLCSR}
 \begin{eqnarray}
\!\!\!\!\!\!Q^2 F^{\gamma\pi}_{\text{LCSR};0}\left(Q^2\right)
\!\!&=&\!\!  N_{\text{T}}\Bigg\{  \int^{\bar{x}_0}_{0} \!\! \bar{\rho}_0(Q^2,x)\frac{dx}{\bar{x}}+
\!\frac{Q^2}{m_{\rho}^2} \int^{1}_{\bar{x}_0}
  \exp\left(
            \Ds \frac{m_{\rho}^2}{M^2}- \frac{Q^2}{M^2}\frac{\bar{x}}{x}
      \right)\bar{\rho}_0(Q^2,x)\frac{dx}{x} + \label{eq:38a} \\
\!\!\!\!\!\!&&\!\!    \left(\frac{\mathbb{A}_{1}(0,s_0;x)}{x}\right)
\underset{x}{\otimes}
  \mathcal{T}^{(1)}(x,y)
\underset{y}{\otimes}
  \psi_0(y)+ \nonumber \\
\!\!\!\!\!\!&&\!\! \frac{Q^2}{m_{\rho}^2}  \int^{1}_{\bar{x}_0}\exp\left(
            \Ds \frac{m_{\rho}^2}{M^2}- \frac{Q^2}{M^2}\frac{\bar{x}}{x}
            \right)\frac{dx}{x}
            \Delta_1(0,\bar{x})
            \mathcal{T}^{(1)}(\bar{x},y) \underset{y}\otimes\psi_0(y)+ O(\mathbb{A}_2)
            \Bigg\}, \label{eq:38b}
\end{eqnarray}
\begin{eqnarray}
\!\!\!\!\!\!Q^2 F^{\gamma\pi}_{\text{LCSR};n}\left(Q^2\right)
\!\!&=&\!\!
\frac{N_\text{T}}{a_s^{\nu_n}(\mu^2)}\Bigg\{
 \left(\frac{\mathbb{A}_{\nu_n}(0,s_0;x)}{x}\right)
\underset{x}{\otimes}\psi_n(x)
  +\left(\frac{\mathbb{A}_{1+{\nu_n}}(0,s_0;x)}{x}\right)
\underset{x}{\otimes}
  \mathcal{T}^{(1)}(x,y)
\underset{y}{\otimes}
  \psi_n(y)\, + \label{eq:38c}\\
\!\!\!\!\!\!&&\!\!\frac{Q^2}{m_{\rho}^2}\int^{1}_{\bar{x}_0} \exp\left(
                                   \Ds \frac{m_{\rho}^2}{M^2}- \frac{Q^2}{M^2}\frac{\bar{x}}{x}
                             \right)
                                   \frac{dx}{x}
            \bigg[ \Delta_{\nu_n}(0,\bar{x})\psi_n(x)+ \Delta_{1+\nu_n}(0,\bar{x})
             \mathcal{T}^{(1)}(\bar{x},y) \underset{y}\otimes\psi_n(y)\bigg] +\nonumber \\
\!\!\!\!\!\!&&          \phantom{
                     \int^{1}_{0}
            \exp(\frac{m_{\rho}^2}{M^2}- \frac{Q^2}{M^2}\frac{\bar{x}}{x})
            }
              + O(\mathbb{A}_2) \Bigg\}\, ,
\label{eq:38d}
\end{eqnarray}
\end{subequations}
%Eqs (38a), (38b), (38c), (38d)
where the functions
$\Delta_{\nu_{n}}(0,\bar{x})$ and $\Delta_{1+\nu_{n}}(0,\bar{x})$
are defined analogously to (\ref{eq:Aeffect-m}) and
(\ref{eq:Aeffect-m=0}), and denote the effective couplings
entering the soft part.
The two equations above represent our final expressions for the TFF
in the dispersive form of RG augmented LCSRs and encapsulate the calibrated
FAPT expansion for this quantity.
Let us also supply some important remarks.
First, the Born contribution in Eq.\ (\ref{eq:38a}) coincides with
the standard one for the zero-harmonic, see Eq.\ (\ref{eq:BornLCSR})
(where $\bar{x}_0\equiv1-x_0=Q^2/(Q^2+s_0)$), while all further
corrections in (\ref{eq:38b}), (\ref{eq:38c}), and (\ref{eq:38d})
appear as the result of the FAPT summation of the radiative corrections
(see Sec.\ \ref{sec:mesonic-FAPT}).
Second, their contributions bear the same minus sign and
for momentum transfers in the range $0.6 \leqslant Q^2 \lesssim 10$ GeV$^2$
have a magnitude a few times smaller than the FOPT results
\cite{Mikhailov:2009kf,Agaev:2010aq}.
The important element of these corrections, i.e., the convolution term
$\mathcal{T}^{(1)}(x,y) \otimes\psi_n(y)$,
is given in App.\ \ref{App:A}.
Third, by significantly reducing the size of the radiation corrections,
the actual constraint for the applicability of the LCSR approach resides
with the high-twist contributions that become important for
$Q^2 \lesssim m^2_\rho$.
%%%%%%%%%%%%%%%%%%%%%%%%%%%%%%%%%%%%%%%%%%%%%%%%%%%%%%%%%%%%%%%%%%%%%%% Figure 3
\begin{figure}[h]
 \centerline{\hspace{0mm}\includegraphics[width=0.65\textwidth]{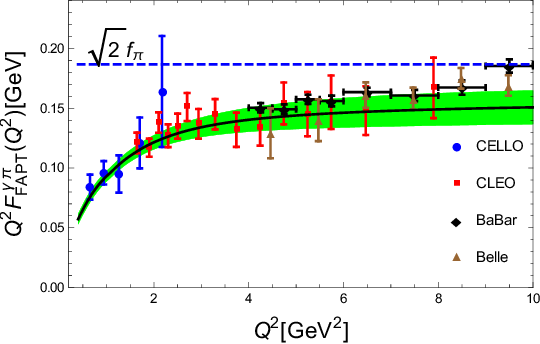}}%Fig2Err.eps
 \caption{\label{fig:pionFF-strips}
    Theoretical predictions for the scaled
    $\gamma^*\gamma\pi^0$ transition form factor
    $Q^2F^{\gamma\pi}_\text{FAPT}(Q^2)$ using the
    BMS DA---black curve---shown in comparison with various
    experimental data up to 10~GeV$^2$ with
    labels as indicated in the figure.
    The green strip around the black curve shows the theoretical uncertainties
    (see text) of the BMS DA obtained with QCD sum rules with nonlocal condensates, see
    \cite{Bakulev:2001pa}.
}
\end{figure}
%%%%%%%%%%%%%%%%%%%%%%%%%%%%%%%%%%%%%%%%%%%%%%%%%%%%%%%%%%%%%%%%%%%%%%%

In order to test the capabilities of the new scheme in more
real terms, we illustrate in Fig.\ \ref{fig:pionFF-strips} its
application to the TFF in comparison with the experimental data.
This version of the figure implements the corrections
done in the included Erratum.
For the sake of definiteness, we employ the family of the bimodal
BMS pion DAs, obtained in \cite{Bakulev:2001pa}.
The corresponding results for the TFF are shown in the form of a
green strip which quantifies the variation of these DAs in terms of their
coefficients $\{1, a_2,  a_4\}$ with the Gegenbauer decomposition
in Eq.\ (\ref{eq:gegen-exp})
(for derivation and justification see
\cite{Bakulev:2001pa,Stefanis:2014yha,Mikhailov:2016klg} and
references therein).

The predictions shown in Fig.\ \ref{fig:pionFF-strips} are obtained by
employing the values of these coefficients at the normalization scale
$\mu^2 \approx 1~\text{GeV}^2$
\cite{Bakulev:2002uc, Bakulev:2004mc} that are given by
$
 \{a_2(\mu^2)
=
 0.20(+0.05/\!-0.06),a_4(\mu^2)
=
 -0.14(+0.09/\!-0.07),\ldots \}$\footnote{
The values $a_2$ and $a_4$ are strongly correlated along the line
$a_2+a_4=$~const.}.
The other LCSR parameters have been fixed by previous investigations to
the values \cite{Khodjamirian:1997tk,Bakulev:2002hk}
$s_0 \approx 1.5~\text{GeV}^2$,
$M^2=0.9$~GeV$^2$, $m_{\rho}^2 \approx 0.6$~GeV$^2$,
$\Lambda_{(4)}^{(1-\text{loop})} \approx 0.3$~GeV , and
$\delta_\text{tw-4}^2(\mu^2)
\approx
 \lambda^2_q/2
\approx
 0.19~\text{GeV}^2$ and are not varied here.

Using Eq.\ (\ref{eq:37}) and the partial
TFF terms $F^{\gamma\pi}_{\text{LCSR};n}$ from
Eqs.\ (\ref{eq:finalFAPTLCSR}), we obtain for
$Q^2F^{\gamma\pi}_\text{FAPT}(Q^2)$ the prediction
shown by the solid black line in Fig.\ \ref{fig:pionFF-strips}.
The (green) strip enveloping this curve indicates the admitted
theoretical variations of the BMS DA in terms of $a_2$ and $a_4$,
while other uncertainties are not considered here.
The interested reader can find estimates of the various
theoretical uncertainties entering the TFF calculation in
\cite{Mikhailov:2016klg,Stefanis:2012yw}.
It is worth noting that the platykurtic pion DA \cite{Stefanis:2014nla}
yields very similar TFF predictions to the BMS-like DAs and has,
therefore, been omitted in Fig.~\ref{fig:pionFF-strips}---\cite{Mikhailov:2016lof}.

The presented prediction is in good agreement with
all experimental data, especially in the low/moderate region of
$Q^2 \leqslant 5$ GeV$^2$, as comparison with the LCSR result
in Fig.\ 4 of \cite{Mikhailov:2016lof} also reveals,
though a more detailed comparison requires additional analysis.
The achieved agreement with the experimental data results from the
decrease of the negative contribution of the resummed radiative
corrections within the applied dispersion representation
together with the twist-six contribution, see the included Erratum.
We restrict ourselves in this work to this qualitative observation,
while a full-fledged analysis of the experimental data within the
presented elaborated FAPT-LCSR approach will be carried out in
future work.
For phenomenological purposes such analysis appears particularly
useful for the expected low-momentum data of the BESIII
Collaboration.

\section{Conclusions}
\label{sec:concl}
In this work we considered the $\pi^0\gamma^*\gamma$ TFF and proposed
an approach, which combines the method of LCSRs based
on dispersion relations, with the renormalization group summation
expressed in terms of the formal solution of the ERBL evolution equation.
We argued that this procedure gives rise to a particular version of FAPT
\cite{Bakulev:2005gw,Bakulev:2006ex} and worked out the technical details.

The advantage of the obtained calibrated FAPT scheme
pertains to process-dependent
boundary conditions imposed on the analytic versions of the couplings
in the deep infrared region in order to preserve the asymptotic behavior
of the TFF prescribed by perturbative QCD.
The resulting theoretical scheme provides the possibility to include
the infinite series of radiative corrections to the considered TFF via RG summation.
The involved techniques are exposed in the text with more
calculational details being given in three dedicated appendices.

Though our focus in this work was primarily on the methodological
aspects of the new framework, we also provided a qualitative
phenomenological application to show the effect on the TFF
of the RG summation of the radiative corrections.
To this end, we presented a qualitative prediction for the scaled
$\pi^0\gamma^*\gamma$ TFF employing BMS-like pion DAs, which we show
in Fig.\ \ref{fig:pionFF-strips} for low to intermediate $Q^2$ values
in comparison with the existing data.
This version of the figure complies with the included Erratum.
We argue that the use of the RG-improved LCSR provides
an improvement relative to the standard LCSR method based on FOPT---especially
at low $Q^2$ \cite{Mikhailov:2016klg}.
Dedicated analysis of the asymptotic regime of the TFF in theory and
in terms of the experimental data will be given in \cite{2018-Ste}.

\acknowledgments
We would like to thank A.\ V.\ Pimikov for useful discussions.
This work was partially supported by the Heisenberg-Landau
Program (Grants 2017 and 2018),
the JINR-BelRFFR grant F16D-004, by the FONDECYT
Postdoctoral Grant No. 3170116,
and by FONDECYT Regular Grant No. 1180344.

\begin{appendix}
\appendix
\section{Next-order perturbation elements}
\label{App:A}
The coefficient function of the partonic subprocess
$\mathcal{T}^{(1)}$ and the evolution kernel $V^{(0)}_+$,
\begin{eqnarray}
  \frac{1}{C_{\rm F}} {\cal T}^{(1)}(x,y)
&=&
  \left[-3 V^{b}  +  g \right]_+(x,y) - 3 \delta(x-y), \label{g} \\
 \frac{1}{C_{\rm F}} V^{(0)}_+(x,y)
&=&
  2\left[{\cal C}\theta(y>x)\frac{x}{y}
         \left(1+\frac{1}{y-x}\right)
         \right]_{+} \equiv 2 \left[V^{a}(x,y) +V^{b}(x,y)\right]_{+}\, ,
\label{eq:V}
\end{eqnarray}
%Eq (A1), (A2)
are determined by the elements
\begin{subequations}
\begin{eqnarray}
  g_{+}(x,y)
&=&
  -2\left[
         \theta(y>x)\frac{\ln\left(1-x/y\right)}{y-x}
         +\theta(y<x)\frac{\ln(1-\bar{x}/\bar{y})}{x-y}
    \right]_{+} \, , \label{eq:g+}  \\
  V^{a}(x,y)
&=&
   {\cal C} \theta(y>x)\frac{x}{y},~~~~~V^{b}(x,y)
=
  {\cal C} \theta(y>x)\frac{x}{y}\left(\frac{1}{y-x}\right)\, ,
\label{eq:Def-va}
\end{eqnarray}
\end{subequations}
%Eq (A3a), (A3b)
where the symbol $\mathcal{C}$ means
$ {\cal C}=\1+\left\{ x \to \bar{x}, y \to \bar{y}\right\} $. \\
The expression for the key convolution term
${\cal T}^{(1)}(x,y)\otimes\psi_n(y)$
in the harmonic expansion can be significantly simplified to get
\begin{eqnarray}
\frac{1}{C_{\rm F}} {\cal T}^{(1)}(x,y)\underset{y}\otimes\psi_0(y)
&=&
    \left[ -3+\frac{\pi^2}{3} - \ln^2\left(\frac{\bar{x}}{x}\right) \right] \psi_0(x) - 2~\psi_0(x), \\
\frac{1}{C_{\rm F}} {\cal T}^{(1)}(x,y)\underset{y}\otimes\psi_n(y)
&=&
    \left[ -3\left(1+v^{b}(n)\right)+\frac{\pi^2}{3} - \ln^2\left(\frac{\bar{x}}{x}\right) \right] \psi_n(x)
    - 2   \sum^n_{l=0,2,\ldots}\!\!\!G_{nl}\psi_l(x) \, ,
\label{eq:barrho-1-b}\\
v^{b}(n)&=& 2\left(\psi(2)-\psi(2+n) \right);~v(n)=1/(n+1)(n+2)-1/2 + 2\left(\psi(2)-\psi(2+n) \right)
\label{eq:vb-v} \, ,
\end{eqnarray}
%Eq (A4), (A5), (A6)
see Appendix A in \cite{Mikhailov:2009kf}.
The quantities $v^{b}(n)$ and $v(n)=-\frac{1}{2 C_{\rm F}}\frac{1}{2}\gamma_0(n)$
are the eigenvalues of the
elements $V^b_+$ and $V^a_+ + V^b_+$ of the one-loop kernel in
Eq.\ (\ref{eq:V}), respectively.
Expression $G_{nl}$ denotes the elements of a calculable triangular matrix
(omitted here) --- see for details \cite{Mikhailov:2009kf,Agaev:2010aq}.

\section{Analytic properties of FAPT couplings}
\label{App:B}
\subsection{Initial FAPT couplings}
\label{subsec:ini-fapt}
In this Appendix we give the expressions for the standard one-loop
running couplings and their FAPT counterparts.
To facilitate the representation, we express them in terms of the
auxiliary variable
$L=\ln(Q^2/\Lambda_\text{QCD}^2)$,
multiplied for simplicity by the term $\beta_0^\nu$.
In other words, we shift the origin of the different coupling images
to the values $a_s \to A_s= \beta_0 a_s =\beta_0 \alpha_s/(4\pi)$,
\begin{subequations}
\label{eq:couplings}
\begin{eqnarray}
A_{s}^{\nu}[L] &=& \frac1{L^\nu}~~~~~~~~~~~~~~~~~~~~~~~~~~~~~~~~~~~~~~~~~~~~~~~~~~~~~~~~~~~~~~~ \mbox{standard pQCD}\, ,
\label{eq:A-s} \\
 {\cal A }^{(1)}_{\nu}[L]
          &=& \frac1{L^{\nu}}
   - \frac{F(e^{-L},1-\nu)}{\Gamma(\nu)} ~~~~~~~~~~~~~~~~~~~~~~~~~~~~~~~~~~~~~~~~~ \mbox{spacelike FAPT}\, ,
 \label{eq:A-F} \label{eq:AU-F} \\
{\mathfrak A}^{(1)}_{\nu}[L_s]
  & = & \frac{\sin\left[(\nu -1)\arccos\left(L_s/\sqrt{(L_s^2+\pi^2)}
                                       \right)
                  \right]}
             {\pi\,(\nu -1) \left(L_s^2+\pi ^2\right)^{(\nu-1)/2}} ~~~~~~~~~~~~~~~~~~ \mbox{timelike FAPT}\, ,
\label{eq:U-F}
\end{eqnarray}
%Eq (B1a), (B1b), (B1c)
where the symbol $[L]$ is used to denote the function argument,
clearly distinguishing it from the $Q^2$ dependence.\footnote{An
expression analogous to (\ref{eq:U-F}) was derived long ago in
\cite{Gorishnii:1983cu,Broadhurst:2000yc} in connection with multiloop
calculations and the Higgs-boson decay into hadrons.}
The spectral density $\rho^{(1)}_\nu$ has the form
($L_\sigma=\ln(\sigma/\Lambda_\text{QCD}^2)$)
\begin{equation}
  \rho_{\nu}^{(1)}[L_\sigma]= \frac{1}{\pi}\textbf{Im}\,\big[a^{\nu}_{(l)}(-\sigma)\big]
= \frac{1}{\pi}\,
  \frac{\sin\left[\nu~\arccos\left(L_{\sigma}/\sqrt{L^2_\sigma+\pi^2}\right)\right] }{\left(L^2_\sigma+\pi^2\right)^{\nu/2}} \, .
\label{eq:spectr-dens}
\end{equation}
\end{subequations}
%Eq (B1d)

Here, $\Ds \frac{F(e^{-L},1-\nu)}{\Gamma(\nu)}$ is the ``pole remover'',
expressed in terms of the Lerch transcendental function
$F(z,s)~(=Li_s(z))$ \cite{Bakulev:2005gw}.
The following equation
\begin{eqnarray}
\label{eq:B2}
  F(z,1-\nu)+\exp\left(i\pi (1-\nu)\right)F\left(1/z,1-\nu\right)
&=&
  \frac{(2 i \pi)^{1-\nu } }{\Gamma(1-\nu)}
  \zeta\left(\nu, \frac{\ln(z)}{2i \pi}\right)
\end{eqnarray}
%Eq (B2)
determines the analytic continuation into the outer region of the
radius of convergence, making use of the Hurwitz zeta function
$\zeta(\nu, z)$.
The first few terms of its asymptotic expansion for
$arg(z)< \pi$ are given by \cite{NIST:DLMF25-11}
\begin{eqnarray}
 \label{eq:A3Hurwitz}
  \zeta(\nu, z)|_{|z|\to \infty}
=
  -\left[\frac{z}{2\pi i}\right]^{1-\nu}\frac{1}{1-\nu}
  +\frac{1}{2}\left[\frac{z}{2\pi i}\right]^{-\nu}+\ldots
\end{eqnarray}
%Eq (B3)
Using this asymptotic expansion for $L \to - \infty$
(which corresponds to $Q^2 \to 0$)
and for $0 \leqslant \nu \leqslant 1$, we obtain for Eq.\ (\ref{eq:B2})
the following expression
\begin{eqnarray}
\label{eq:B4}
&&  F(e^{|L|},1-\nu)+ O\left(e^{-|L|}\right)
=
  \frac{(2 \pi i)^{1-\nu } }{\Gamma(1-\nu)}
  \zeta\left(\nu, \frac{|L|}{2i \pi}\right)
=
  - \frac{1}{\Gamma(2-\nu)}\left[|L|\right]^{1-\nu}
  + \frac{i\pi}{\Gamma(1-\nu)}\left[|L|\right]^{-\nu}+ \ldots \; .
\end{eqnarray}
%Eq (B4)
By substituting the asymptotic form of this equation into
(\ref{eq:A-F}), we get
\begin{eqnarray}
&& {\cal A }^{(1)}_{\nu}[L] \stackrel{L \to -\infty}{\Longrightarrow}
   \frac1{L^{\nu}}
   + \frac{1}{\Gamma(\nu)\Gamma(2-\nu)}|L|^{1-\nu}
   -\frac{i\pi}{\Gamma(\nu)\Gamma(1-\nu)}|L|^{-\nu}+ \ldots \, ,
\label{eq:B5}
\end{eqnarray}
%Eq (B5)
where the second term leads to a divergence
in the limit $\sim |L|^{1-\nu}$ when $L \to -\infty$.
Employing Eqs.\ (\ref{eq:B5}) and (\ref{eq:U-F}), one can
then obtain the range of values of the functions
${\cal A }^{(1)}_{ \nu },~{\mathfrak A}^{(1)}_{\nu}$, notably,
\begin{subequations}
\label{eq:B6}
\begin{eqnarray}
{\cal A }^{(1)}_{0}[L]=1;&~{\cal A }^{(1)}_{0<\nu <1}[L \to -\infty] \rightarrow |L|^{1-\nu};&~
{\cal A }^{(1)}_{1}[L \to -\infty]\to 1;~{\cal A }^{(1)}_{\nu >1}[L \to -\infty]\to 0\, ,
\label{eq:B6a} \\
                         &                                                                   & \nonumber \\
{\mathfrak A}^{(1)}_{0}[L]=1;&~{\mathfrak A}^{(1)}_{0<\nu <1}[L \to -\infty]
\rightarrow \left(\sqrt{L^2+\pi^2}\right)^{1-\nu};&~{\mathfrak A}^{(1)}_{1}[L
\to -\infty]\to 1;~{\mathfrak A}^{(1)}_{\nu >1}[L \to -\infty]\to 0\,.
\label{eq:B6b}
\end{eqnarray}
\end{subequations}
%Eq (B6a), (B6b)

One observes that the couplings
${\cal A }^{(1)}_{ \nu }[L], ~{\mathfrak A}^{(1)}_{ \nu }[L]$
become unbounded in the vicinity of $Q^2=0$ and $0< \nu \leqslant 1$.
In some sense, the well-known singularity of the standard
running coupling $A^{\nu}_{s}[L]$ in Eq.\ (\ref{eq:A-s}) for
$L=0$ and $\nu> 0 $ turns in the limit $L \to -\infty$
into a singularity of the FAPT couplings for
$0< \nu <1$, cf.\ (\ref{eq:A-F}) and (\ref{eq:U-F}).
Therefore, these results for
${\cal A }^{(1)}_{\nu}[L], ~{\mathfrak A}^{(1)}_{ \nu }[L ]$
cannot be directly used in the vicinity of $Q^2=0$
and $0< \nu \leqslant 1$,
where the FAPT couplings are ill-defined.
One needs to ``calibrate'' these couplings in this regime
by demanding that they vanish.
This intervention guarantees at the same time that observables
calculated with them, e.g., the TFF, have the correct UV asymptotics
following from pQCD.
To this end, we redefine the couplings
${\cal A }^{(1)}_{\nu},~{\mathfrak A}^{(1)}_{\nu}$
in the vicinity of $Q^2=0$ and for $0< \nu \leqslant 1$, to be
${\cal A }^{(1)}_{\nu}[-\infty]=0,~{\mathfrak A}^{(1)}_{\nu}[-\infty]=0$,
while the behavior of these couplings for $\nu > 1$ remains unaffected.

\subsection{Generalized FAPT coupling $\bm{\I_{\nu}}$}
\label{subsec:gen-fapt}
\begin{figure}[htb]
%\begin{minipage}[b]{.45\linewidth}
\includegraphics[width=.6\linewidth]{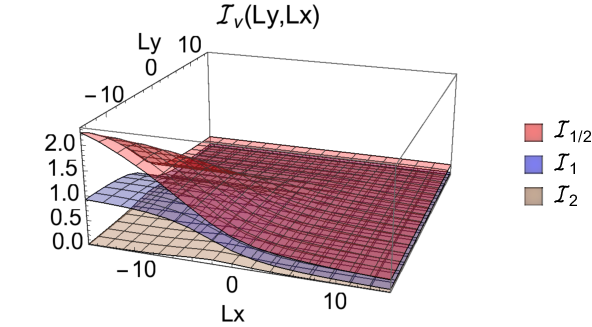}%plotILog3D.eps
%\end{minipage}
\vspace{-0.3cm}
\caption{
         3D Plot of the generalized coupling
         $\I_{\nu}(\ln(x),\ln(y))$ using logarithmic scales.
         The abbreviations in the plot mean
         $Lx=\ln(x),~Ly=\ln(y)$, with $\I_{\nu}$ being considered for three
         different values of the index $\nu=1/2,1,2$ and $N_f=3$.
        \label{fig6}}
\end{figure}

\begin{figure}[htb]
%\begin{minipage}[b]{.45\linewidth}
\includegraphics[width=.415\linewidth]{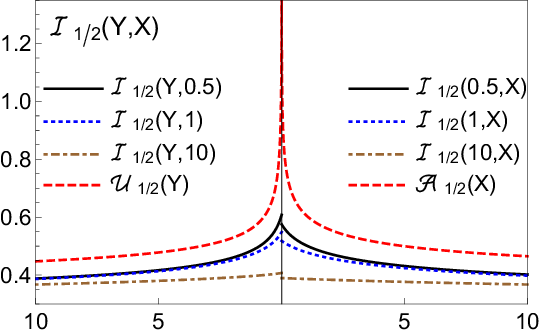}
%\end{minipage}
~\includegraphics[width=.415\linewidth]{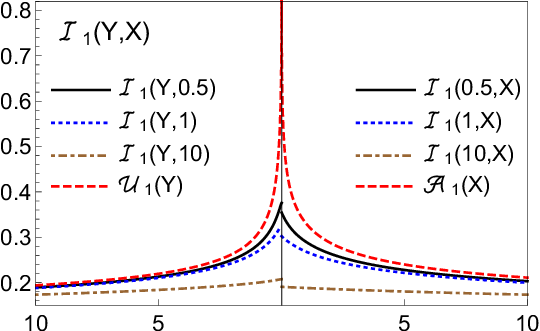}
~\includegraphics[width=.415\linewidth]{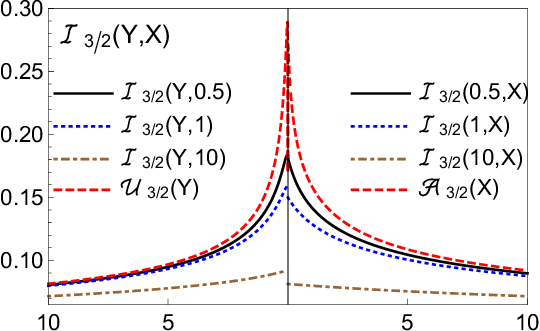}
~\includegraphics[width=.415\linewidth]{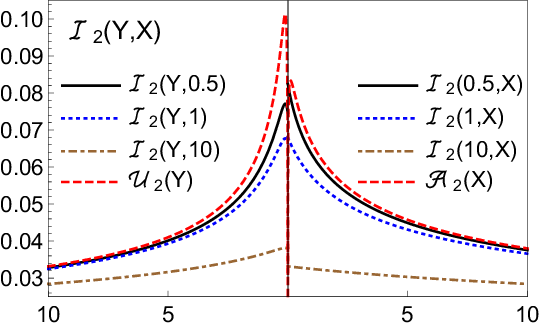}
\vspace{-0.3cm}
\caption{``Distorting mirror'' symmetry in the 2D projections of
         the 3D plots of $\I_{\nu}$.
         The couplings $\I_{\nu}(y,\text{fixed})$, $\I_{\nu}(\text{fixed},x)$
         are taken at different values of the index $\nu=1/2,1,3/2,2$.
\label{fig7}
}
\end{figure}

During the calculation considered in Subsec.\
\ref{subsec:rad-corr-disp-rel}, a new, more general,
two-argument coupling $ \I_{\nu}(y,x)$ appeared ``naturally'' in
Eq.\ (\ref{eq:4}), viz.,
\begin{eqnarray}
 \label{eq:B7}\sqrt{}
  \I_{\nu}(y,x)=
\int_{y}^\infty \frac{ds}{s+x} \rho_{\nu}(s)
&=&\left\{
\begin{array}{l}
 \Ds
  \left[\M_{\nu}(y) -~~ x\int_{y}^\infty \frac{ds}{s(s+x)} \rho_{\nu}(s) \right]
  \leqslant \M_{\nu}(y),~\text{for}~ \rho_{\nu}\geqslant 0\, ,
\label{eq:A-def} \\
\Ds \left[\A_{\nu}(x)\! -\! \int_{0}^y \frac{ds}{s+x} \rho_{\nu}(s) \right]
   \leqslant \A_{\nu}(x), ~\text{for}~ \rho_{\nu}\geqslant 0\, ,
 \end{array}
 \right. \\
  \I_{\nu}(y,x \to 0)
&=&\M_{\nu}(y),
~~\I_{\nu}(y \to 0, x)
=\A_{\nu}(x),
~~\I_{1}(y \to 0,x \to 0)=  \M_{1}(0)=\A_{1}(0)\,.
\label{eq:B8}
\end{eqnarray}
%Eq (B7), (B8)
The coupling $\I_{\nu}(y,x)$ is regular for $y>0, x>0$,
while for $y=0 $ or $x=0$ it reduces to the initial FAPT couplings
in accordance with Eq.\ (\ref{eq:B8}).
The behavior of $\I_{\nu}(y,x)$ with respect to the arguments $(x,y)$
is illustrated in Fig.\ \ref{fig6}, while the graphics showing its
behavior when one of its arguments is fixed is displayed in
Fig.\ \ref{fig7}.
These figures are corrected versions of the published
ones after removing an error in the code---see the Erratum below.
One appreciates in this figure the smallness of the next-order
FAPT coupling $\I_{2}$.
%%%%%%%%%%%%%%%%%%%%%%%%%%%%%%%%%%%%%%%%%%%%%%%%%%%%%%%%%%%%%%%%%%%%%%%%%%%%%%% Figure 4
%%%%%%%%%%%%%%%%%%%%%%%%%%%%%%%%%%%%%%%%%%%%%%%%%%%%%%%%%%%%%%%%%%%%%%%%%%%%%%%

%%%%%%%%%%%%%%%%%%%%%%%%%%%%%%%%%%%%%%%%%%%%%%%%%%%%%%%%%%%%%%%%%%%%%%%%%%%%%%% Figure 5

\section{Two scales effective coupling}
\label{App:C}
Here we outline the calculation of the key element of the dispersion
integral discussed in Sec.\ \ref{subsec:h-LCSR}, viz.,
\begin{eqnarray}
\label{eq:B-H-fapt}
N_{\text{T}}  \int_{s_0}^{\infty}
        \frac{ds}{s+q^2}\rho(Q^2,s)
=
  F^{\gamma^\ast\pi}_\text{FAPT}(Q^2;m^2,s_0)\, ,
\end{eqnarray}
%Eq (C1)
at the low limit $s_0$.
After changing the integration variable to
$ s \to\sigma= -(-s\bar{y}+Q^2y) \geq 0 $, the low limit becomes
$s(y)=s_0\bar{y}-Q^2y$.
The value of the low limit $s(y)> m^2$ of this function
leads to a new constraint for the range of integration in the variable
$\sigma$, notably, $\sigma> s(y)$.
On the other hand, if $s(y)\leqslant m^2$,
one should start to integrate with $\sigma=m^2$,
where $\rho_{\nu}(\sigma)\neq 0 $.
In the range $s_0>m^2$, one then obtains
\begin{eqnarray}
  I_n(Q^2,q^2)
&=&
 - \theta(s_0 > m^2)\left[\theta(s(y)> m^2)\int_{s(y)}^\infty ds
 \frac{\rho_{\nu_n}(\sigma)}{\sigma(\sigma+Q(y))}
  +\theta(s(y)\leqslant m^2)\int_{m^2}^\infty ds
  \frac{\rho_{\nu_n}(\sigma)}{\sigma(\sigma+Q(y))}
  \right]\, ,
\label{eq:Idecompose-1}
\end{eqnarray}
%Eq (C2)
while for the case $s_0 < m^2$ only the second term survives,
i.e.,
\begin{eqnarray}
  I_n(Q^2,q^2)
&=&
  -\theta(s_0 < m^2)\int_{m^2}^\infty ds
  \frac{\rho_{\nu_n}(\sigma)}{\sigma(\sigma+Q(y))}\, .
\label{eq:Idecompose-2}
\end{eqnarray}
%Eq (C3)
Substituting Eq.\ (\ref{eq:4a}) into
Eq.\ (\ref{eq:Idecompose-1}), one arrives at
the final expression for $I_n$,
\begin{eqnarray}
  I_n(Q^2,q^2)
&=&
  T_0(Q^2,q^2;y)\bigg\{ \theta\left(y <y_0\right)
                      \Big[\I_{\nu_n}(s(y),Q(y))-\M_{\nu_n}(s(y))\Big]
\nonumber \\
&&\phantom{ T_0(Q^2,q^2;y)\Big\{ }\!\!\!
                      +\theta\left(y \geqslant y_0\right)
                      \Big[\I_{\nu_n}(m^2,Q(y))-\M_{\nu_n}( m^2)\Big]
                \bigg\} ,
\label{eq:C4}
\end{eqnarray}
%Eq (C4)
where $y_0=(s_0-m^2)/(s_0+Q^2)$
and the couplings $\mathcal{A}_\nu$ and $\mathfrak{A}_\nu$
appear as the limits of $\I_{\nu}$ by allowing their arguments
to approach zero, cf.\ (\ref{eq:4d}).
The effective coupling $\mathbb{A}_\nu(m^2,s_0;y)$, following from
Eqs.\ (\ref{eq:Idecompose-1}), (\ref{eq:Idecompose-2}), reads
\begin{eqnarray}
\label{eq:eff-coupl2-C5}
\! \mathbb{A}_\nu(m^2,s_0;y)
&=&\!\!\left\{
\begin{array}{l}
 \Ds \theta\left(y <y_0\right) \Big[\I_{\nu_n}(s(y),Q(y))-\M_{\nu_n}(s(y))\Big] +
  \theta\left(y\geqslant y_0 \right)\Big[\I_{\nu_n}(m^2,Q(y))-\M_{\nu_n}( m^2)\Big]
\,,~ m^2 < s_0 \\
 \\
\I_{\nu_n}(m^2,Q(y))-\M_{\nu_n}( m^2)\,,~ m^2 >s_0 \, ,
\end{array}
 \right.
\end{eqnarray}
%Eq (C5)
whereas the quantity $\mathbb{A}_\nu(m^2,y)$ in the vicinity of
$y_0$ for $s(y_0)=m^2$ is a continuous function as it
follows from the properties expressed in (\ref{eq:4d}).
In the limit $s_0,m^2 \to 0$, one has
$\mathbb{A}_\nu(0,y) \to \left[\A_{\nu}(Q(y))-\A_\nu( 0)\right]$,
which completes the argument.
\end{appendix}

\section*{ERRATUM}
\begin{figure}[hbt] %\unitlength=1mm
%\centering\epsfig{file=mb(mu).eps,width=8.cm}
\includegraphics[width=.45\linewidth]{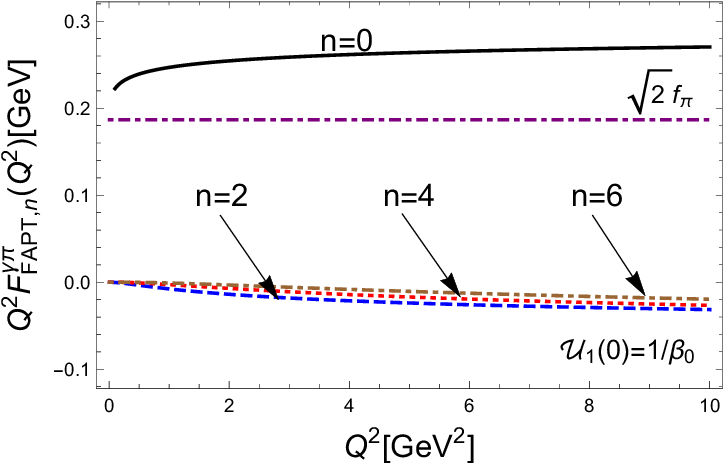}
~~~~~\includegraphics[width=.45\linewidth]{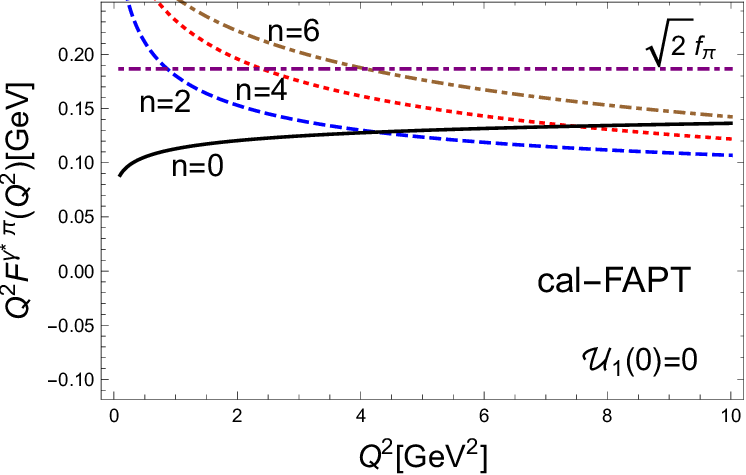}
%\vspace{-0.5cm}
\caption{Corrected graphical results of the right panels of Figs./ 1
(left) and 2 (right) of the journal version.
The curves in the left panel were calculated with the modified
FAPT scheme, $Q^2 F^{\gamma\pi}_{\text{FAPT},n}(Q^2)$, according to Eq.\ (22) for
${\mathfrak A}^{(1)}_{1}(0)=1/\beta_0$ from (20b), and with
$m^2 =4 m_\pi^2 \approx 0.08$ GeV$^2$ from Eq.\ (20a) for the higher harmonics.
Those in the right panel follow from Eq.\ (22)
in cal-FAPT with $\M_{\nu}(0)=\A_{\nu}(0)=0$ according to Eq.\ (20b).
\label{fig1-Err}}
\end{figure}

The graphics shown in the right panels of Figs.\ 1 and 2 of the
journal version are the result of an erroneous computer code.
The corresponding corrected curves are displayed here in Fig.\ \ref{fig1-Err}.

%%%%%%%%%%%%%%%%%%%%%%%%%%%%%%%%%%%%%%%%%%%%%%%%%%%%%%%%%%%%%%%%%%%%%%%%%%%%%%%%%%%%%%%%%%%%%%%%%%% Errarum Figure 1; here Figure 6
%%%%%%%%%%%%%%%%%%%%%%%%%%%%%%%%%%%%%%%%%%%%%%%%%%%%%%%%%%%%%%%%%%%%%%%%%%%%%%%%%%%%%%%%%%%%%%%%%%%

Analogously, also the predictions shown in Fig.\ 3 in the journal are
wrong.
The correct contribution to the TFF, comprising the twist-two and
twist-four terms, is shown as a dashed line in Fig.\ \ref{fig2-Err} below.

As one sees, FAPT resummation of the radiative corrections induces a sizeable
suppression of the transition form factor.
To get agreement with the experimental data, it is mandatory to
include into expression (36) the twist-six contribution.
This is done by taking into account the spectral density
$\bar{\rho}(Q^{2}\!,x)$ for the initial LCSR in Eq.\ (28), the
corresponding contribution $\bar{\rho}_\text{tw-6}$
\cite{Agaev:2010aq,Mikhailov:2016klg}, given by
\begin{eqnarray}
 \label{eq:tw-6}
    \bar{\rho}_\text{tw-6}(Q^{2}\!,x)
=
    8\pi \frac{C_F}{N_c}
    \frac{ \alpha_s\langle\bar{q} q\rangle^2}{f_\pi^2}\frac{x}{Q^4}
    \left[
        \!-\!
        \left(\frac{1}{1-x}\right)_+
        \!+\!\left(2\delta(\bar{x})-4 x\right)\!+\!
        \left(
         3x+2x\log{x}
        \!+\!
        2x\log{}\bar{x}
        \right)
    \right] \, ,
\end{eqnarray}
%Eq (19)
where
$\alpha_s=0.494$ and $
 \langle \bar{q} q\rangle^2
=
 \left(0.240 \pm 0.01 \right)^6$ GeV$^6$ \cite{Agaev:2010aq}.
The resulting TFF is shown in the figure in terms of a solid line.
Thus, the conclusion drawn in the journal version that one gets
agreement with the data by taking into account only the twist-two and
twist-four terms has to be corrected accordingly.
%%%%%%%%%%%%%%%%%%%%%%%%%%%%%%%%%%%%%%%%%%%%%%%%%%%%%%%%%%%%%%%%%%%%%%% Erratum Figure 2; here Figure 7
\begin{figure}[h]
 \centerline{\hspace{0mm}\includegraphics[width=0.65\textwidth]{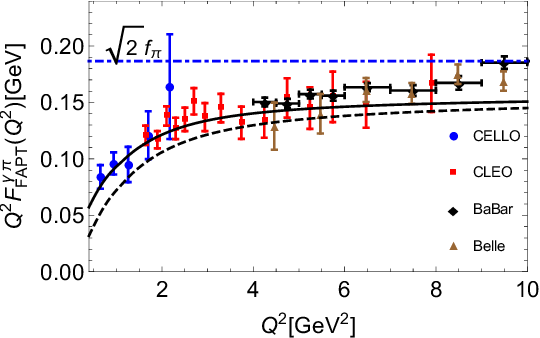}}
 \caption{\label{fig2-Err}
    Theoretical prediction from Eq.\ (38) for the scaled
    $\gamma^*\gamma\pi^0$ transition form factor
    $Q^2F^{\gamma\pi}_\text{FAPT}(Q^2)$ using the
    BMS DA \cite{Bakulev:2001pa}---dashed line---shown in comparison with various
    experimental data up to 10~GeV$^2$ with
    labels as indicated in the figure.
    The solid line shows the analogous theoretical prediction obtained
    by including into the TFF the twist-six contribution.}
\end{figure}
%%%%%%%%%%%%%%%%%%%%%%%%%%%%%%%%%%%%%%%%%%%%%%%%%%%%%%%%%%%%%%%%%%%%%%%

The deviations of the results given in Figs.\ 4 and 5 of the journal
version differ from their corrected counterparts by less than about 5 percent
and are therefore not shown here.

\acknowledgments
We would like to thank Alex Pimikov for help with the numerical
computations.

\end{document}